\documentclass[sigconf]{acmart}

\newcommand{\QUOTE}[1]{\textsf{\emph{``#1''}}}

\newcommand{\PROMPT}[1]{%
    \begingroup
    \spaceskip=0.3em plus 0.1em minus 0.1em 
    \textsf{\small{\fontfamily{cmss}\selectfont ``#1''}}%
    \endgroup
}
\usepackage{xcolor}

\usepackage{comment}
\usepackage{ifthen}
\usepackage{xspace}
\usepackage{etoolbox}
\usepackage{multirow}
\usepackage{makecell}
\usepackage{enumitem}

\usepackage{graphicx}

\AtBeginDocument{%
  }

\copyrightyear{2026}
\acmYear{2026}
\setcopyright{cc}
\setcctype{by}
\acmConference[CHI '26]{Proceedings of the 2026 CHI Conference on Human Factors in Computing Systems}{April 13--17, 2026}{Barcelona, Spain}
\acmBooktitle{Proceedings of the 2026 CHI Conference on Human Factors in Computing Systems (CHI '26), April 13--17, 2026, Barcelona, Spain}
\acmDOI{10.1145/3772318.3791308}
\acmISBN{979-8-4007-2278-3/2026/04}

\def\name{TouchScribe}

\setcopyright{none}

\begin{document}

\title{TouchScribe: Augmenting Non-Visual Hand-Object Interactions with Automated Live Visual Descriptions}

\author{Ruei-Che Chang}
\email{rueiche@umich.edu}
\affiliation{
\department{Computer Science and Engineering}
 \institution{University of Michigan}
 \city{Ann Arbor, MI}
 \country{USA}
}
\author{Rosiana Natalie}
\email{rosianan@umich.edu}
\affiliation{
\department{Michigan Institute for Data \& AI in Society}
 \institution{University of Michigan}
 \city{Ann Arbor, MI}
 \country{USA}
}
\author{Wenqian Xu}
\email{wxtu@umich.edu}
\affiliation{
 \institution{University of Michigan}
 \city{Ann Arbor, MI}
 \country{USA}
}
\author{Jovan Zheng Feng Yap}
\email{jovanyap@umich.edu}
\affiliation{
\department{Computer Science and Engineering}
 \institution{University of Michigan}
 \city{Ann Arbor, MI}
 \country{USA}
}
\author{Tiange Luo}
\email{tiangel@umich.edu}
\affiliation{
 \institution{University of Michigan}
 \city{Ann Arbor, MI}
 \country{USA}
}
\author{Venkatesh Potluri}
\email{potluriv@umich.edu}
\affiliation{
\department{School of Information}
 \institution{University of Michigan}
 \city{Ann Arbor, MI}
 \country{USA}
}
\author{Anhong Guo}
\email{anhong@umich.edu}
\affiliation{
\department{Computer Science and Engineering}
 \institution{University of Michigan}
 \city{Ann Arbor, MI}
 \country{USA}
}

\begin{abstract}
People who are blind or have low vision regularly use their hands to interact with the physical world to gain access to objects' shape, size, weight, and texture. However, many rich visual features remain inaccessible through touch alone, making it difficult to distinguish similar objects, interpret visual affordances, and form a complete understanding of objects. In this work, we present TouchScribe, a system that augments hand-object interactions with automated live visual descriptions. We trained a custom egocentric hand interaction model to recognize both common gestures (e.g., grab to inspect, hold side-by-side to compare) and unique ones by blind people (e.g., point to explore color, or swipe to read available texts). Furthermore, TouchScribe provides real-time and adaptive feedback based on hand movement, from hand interaction states, to object labels, and to visual details. Our user study and technical evaluations demonstrate that TouchScribe can provide rich and useful descriptions to support object understanding. Finally, we discuss the implications of making live visual descriptions responsive to users' physical reach.
\end{abstract}



% \settopmatter{printfolios=true}

\begin{teaserfigure}
  \vspace{-1pc}
  \includegraphics[width=\textwidth, alt={The figure is divided into six panels (a–f) showing how TouchScribe helps blind and visually impaired users explore spice bottles. Panel (a) shows a person sitting at a kitchen counter reaching for spice bottles; the caption reads “Users explore spice bottles using TouchScribe.” Panel (b) shows a close-up of someone holding a red spice bottle, with a teal bottle in the background. The caption says “When holding the object, users get hierarchical feedback,” and TouchScribe provides three levels of spoken feedback: “I see your left hand is holding,” “You are grabbing a red spice bottle,” and “The bottle is a Trader Joe’s Chile Lime Seasoning Blend with a bright red label and bold white and green lettering that reads ‘Just the right amount of salt and heat – Net Wt. 2.9 oz (82g).’” Panel (c) shows a person holding the red bottle, with the caption “TouchScribe allows speech query.” The user asks, “How many calories does it have?” and TouchScribe replies, “It has 0 calories per serving.” Panel (d) shows a hand holding the bottle while another finger points to the label. The caption says “When using discrete hold+pointer,” and the system provides feedback labeled as “4 color label as finger moves,” listing: white, red, red, green, green, etc. Panel (e) shows a hand swiping upward on the bottle while holding it. The caption says “When using discrete hold+swipe-up,” and TouchScribe provides “5 text label reads from top to bottom,” including: “TRADER JOE’S CHILE LIME SEASONING BLEND,” “JUST THE RIGHT AMOUNT OF SALT AND HEAT,” and “NET WT. 2.9 OZ (82g).” Panel (f) shows a person holding two spice bottles, one in each hand: a red bottle in the right hand and a green one in the left. The caption reads “When holding similar objects, they get comparisons.” TouchScribe says, “Your right hand holds a red bottle, and your left holds a green one.” It then provides “6 visual comparison.” Similarities: both are Trader Joe’s seasonings in clear glass bottles. Differences: the Chile Lime bottle in the left hand is red with bold text, while the Oregano bottle in the right hand is green with smaller text.}]{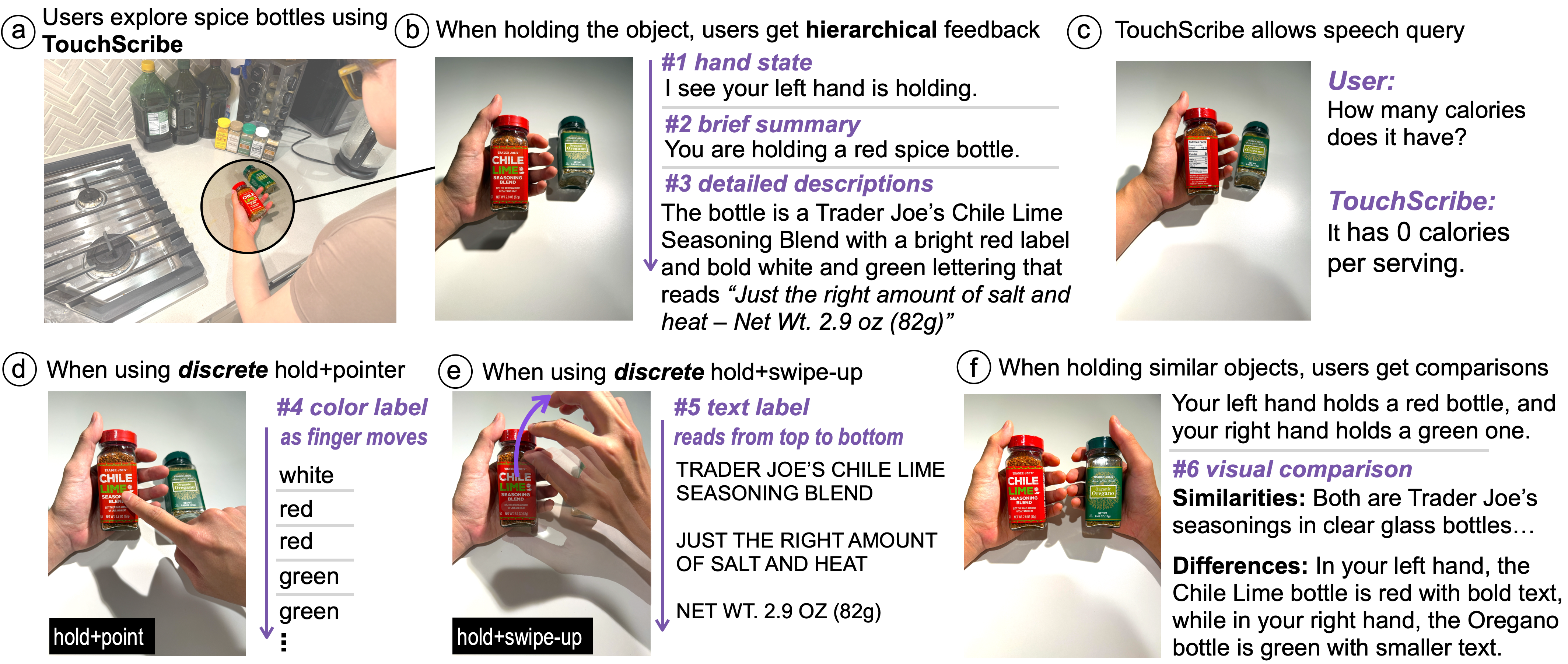}
  \vspace{-1pc}
  \caption{TouchScribe augments hand-object interactions with live visual descriptions.
  (a) BLV users use TouchScribe to explore objects placed on the shared kitchen counter.
  (b) While holding an object, BLV users receive hierarchical descriptions, starting with the hand state, followed by a brief summary, and then detailed descriptions.
  (c) They can also ask TouchScribe for visual information using speech.
  (d) Through finger gestures, users can access object details; for example, holding the object and pointing with the other hand reads its color; (e) holding and swiping up reads the text.
  (f) When holding two objects simultaneously, users receive a visual comparison highlighting their similarities and differences.
  }
  \Description{The figure is divided into six panels (a–f) showing how TouchScribe helps blind and visually impaired users explore spice bottles. Panel (a) shows a person sitting at a kitchen counter reaching for spice bottles; the caption reads “Users explore spice bottles using TouchScribe.” Panel (b) shows a close-up of someone holding a red spice bottle, with a teal bottle in the background. The caption says “When holding the object, users get hierarchical feedback,” and TouchScribe provides three levels of spoken feedback: “I see your left hand is holding,” “You are grabbing a red spice bottle,” and “The bottle is a Trader Joe’s Chile Lime Seasoning Blend with a bright red label and bold white and green lettering that reads ‘Just the right amount of salt and heat – Net Wt. 2.9 oz (82g).’” Panel (c) shows a person holding the red bottle, with the caption “TouchScribe allows speech query.” The user asks, “How many calories does it have?” and TouchScribe replies, “It has 0 calories per serving.” Panel (d) shows a hand holding the bottle while another finger points to the label. The caption says “When using discrete hold+pointer,” and the system provides feedback labeled as “4 color label as finger moves,” listing: white, red, red, green, green, etc. Panel (e) shows a hand swiping upward on the bottle while holding it. The caption says “When using discrete hold+swipe-up,” and TouchScribe provides “5 text label reads from top to bottom,” including: “TRADER JOE’S CHILE LIME SEASONING BLEND,” “JUST THE RIGHT AMOUNT OF SALT AND HEAT,” and “NET WT. 2.9 OZ (82g).” Panel (f) shows a person holding two spice bottles, one in each hand: a red bottle in the right hand and a green one in the left. The caption reads “When holding similar objects, they get comparisons.” TouchScribe says, “Your right hand holds a red bottle, and your left holds a green one.” It then provides “6 visual comparison.” Similarities: both are Trader Joe’s seasonings in clear glass bottles. Differences: the Chile Lime bottle in the left hand is red with bold text, while the Oregano bottle in the right hand is green with smaller text.}
  \label{fig:teaser}
  \vspace{0pc}
\end{teaserfigure}

\maketitle 
\begin{table*}[t]
\centering
\vspace{-0.5pc}
\caption{Overview of research and commercial apps for providing live visual descriptions leveraging different information cursors for real-world understanding.}
\vspace{-0.5pc}
\begin{tabular}{|>{\raggedright\arraybackslash}m{4.5cm}|
                >{\raggedright\arraybackslash}m{2.8cm}|
                >{\raggedright\arraybackslash}m{6cm}|}
\hline
\textbf{Assistive System} & \textbf{Cursor Type} & \textbf{Information Type} \\ \hline
Orcam~\cite{Orcam}, FingerReader~\cite{FingerReader2}, VizLens~\cite{VizLens} and StateLens~\cite{StateLens} & Finger tip & Text \\ \hline
Medeiros et al.~\cite{Medeiros2017} \newline Stearns et al.~\cite{FingerClothing} & Finger tip & Clothing color and texture \\ \hline
EyeRing~\cite{EyeRing} & Finger tip & Barcode, currency \\ \hline
SeeingAI~\cite{seeingai} & Camera motion & Human, currency, barcode, object, color, lightness, and text \\ \hline
WorldScribe~\cite{worldscribe} & Camera motion & Object labels, general and detailed descriptions \\ \hline
\textbf{TouchScribe (this work)} & \textbf{Hands and fingers} & \textbf{Hand states, color, text, brief and detailed object descriptions, and object comparison} \\ \hline
\end{tabular}
\vspace{0pc}
\label{tab:cursors}
\end{table*}

\section{Introduction}
People who are blind or have low vision (BLV) often rely on touch to explore and interact with the physical world, using their hands to perceive essential attributes of objects such as shape, size, weight, and texture~\cite{withagen2010tactile, withagen2012haptic, leo2022early}. 
However, many rich visual features remain inaccessible through touch alone, making it difficult to fully understand an object's appearance and functionality.
For example, it is difficult to distinguish between grocery store products with similar shapes but different colors, patterns, or surface details without visual cues~\cite{Lee2021Grocery, Lee2020Shoppers}. 
Certain visual affordances, such as identically shaped seasoning bottles from different brands, similar tea bags with or without caffeine, or small printed ingredient labels, may be entirely imperceptible through touch.

Currently, BLV individuals can capture photos and engage in conversations with AI systems to obtain visual descriptions~\cite{bemyai, seeingai, envisionai}, or receive live narration to explore their surroundings~\cite{worldscribe, chatGPTVideo}.
However, these AI systems often struggle to accurately identify the specific object of interest within an image (as noted in~\cite{lee2021leveraging, HandsHoldingClues}), generate long-form descriptions that contain unnecessary information, and their turn-taking interaction style can impede the quick retrieval of desired visual information.
In contrast, hands provide a natural and intuitive interface for interacting with the physical world~\cite{lee2021leveraging, lee2020hand, Lee2019ASSETS, HandsHoldingClues}. 
Their movements are indicative of a person’s intent~\cite{bolt1980put} and objects of interest~\cite{cai2018understanding, Zhou2022UIST}, making them integral to how BLV individuals access and understand their environment through tactile exploration~\cite{goldreich2003tactile, TactileFixations}.
Hence, in this paper, we investigate the question: \emph{How can natural hand interactions be leveraged to support BLV people to access rich visual information about objects of interest?}

To address this question, we introduce {\name}, a system that provides live visual descriptions driven by hand-object interactions. 
{\name} supports a set of hand gestures inspired by prior research on discreet gestures preferred by BLV individuals~\cite{Oh2017, Oh2014}, such as pointing to an object held in the other hand to read its color~\cite{FingerClothing, Medeiros2017} or swiping across it to access available text (Figure~\ref{fig:teaser}d, e).
{\name} also incorporates common gestures used with sight, such as touching or holding objects to signal interest (Figure~\ref{fig:teaser}b), and comparing two items side-by-side to explore visual similarities or differences (Figure~\ref{fig:teaser}f). 
Based on different hand interactions with objects, {\name} provides hierarchical feedback, including hand states for users to confirm that hand events are correctly identified, a brief object overview, and rich visual details (Figure~\ref{fig:teaser}b). 

{\name} was prototyped using a neck mount with an attached smartphone (Figure~\ref{fig:setup}). It detects fine-grained hand-object interactions, such as when both hands engage the same or different objects, or when an object is flipped, to deliver adaptive feedback that aligns with the user’s evolving focus and intent. 
To support these hand-object interactions, we fine-tuned a custom egocentric hand gesture recognition model that interprets different hand gestures as information cursors to identify objects or information of interest. 
The underlying gesture recognition model is lightweight enough to run on live video feeds, and the wide camera field of view (FoV) provides broad coverage, though at the cost of accuracy due to inherent model limitations and distortion from the wide-angle lens.
{\name} addresses these by integrating smoothing algorithms to mitigate intermittent recognition and extract keyframes. It also supports visual question answering (VQA), enabling users to freely query visual details when needed (Figure~\ref{fig:teaser}c).

We conducted a study with eight BLV participants to collect both qualitative and quantitative data, aiming to understand their experiences with using {\name} across various object-understanding tasks in our lab-controlled environment, and to evaluate the accuracy and latency of descriptions.
Through qualitative analysis, we found that participants generally perceived {\name} interactions as intuitive (M=5.63 out of 7), with the provided descriptions being accurate (M=5.5) and comprehensive (M=6.5). Participants also felt a sense of control in the descriptions for object understanding (M=5.13).
However, participants reported moderate cognitive effort (as measured by NASA-TLX) and a noticeable learning curve in hand positioning and gesture recognition with camera-enabled assistive technologies (ATs). 
Also, through our technical evaluation, we reported quantitative results to reflect {\name}'s performance in our user study, including the accuracy of our custom hand posture recognition model in the live stream ($F_1=0.77$), the latency between detected hand movements and different types of descriptions (from 0.56s to 14s), and the accuracy of the descriptions (from 67.83\% to 93.27\%).

Through the study, we identified several gaps in the current {\name} prototype that limit its practical use. 
For example, while the wide camera FoV offered broader coverage, it also introduced inaccuracies due to image distortion. In addition, interpreting the intent behind diverse natural hand–object interactions remained challenging, and at times the system produced information overload during rapid hand gesture changes.
Based on these findings, we discuss the implications to make {\name} generalizable for broader real-world situations in the future, such as customizations to different gesture preferences, integrating haptic-audio feedback for camera aiming, leveraging other gesture and object recognition techniques to improve accuracy, and making live visual descriptions responsive to users' further physical reach.

\vspace{1pc}
In summary, our work contributes:
\begin{itemize}
% \vspace{-0.3pc}
\item[\emph{(i)}] {\name}, a novel prototype system that generates live, rich object descriptions based on multiple hand-object interactions, moving beyond the single interaction and information types supported in earlier systems (Table~\ref{tab:cursors}).
\item[\emph{(ii)}] A user study and technical evaluation demonstrating the intuitiveness of {\name}, usefulness of its descriptions, and its overall user experience.
\item[\emph{(iii)}] Lessons learned from the development and evaluation of {\name}, and design implications for employing egocentric camera-enabled, real-time assistive technologies in the real world.
\end{itemize}

\section{Related Work}
Our work builds upon and connects three key research domains. 
First, prior research on hand-based interactions has shown that the expressive and intentional nature of hands provides a compelling alternative to device-based input (e.g., controllers) though user preferences vary depending on different contexts, which informed our selection of gestures in {\name}. 
Second, studies on the use and limitations of hand interactions in current ATs for accessing real-world information revealed opportunities for {\name} to incorporate more expressive hand–object interactions and deliver richer visual information. 
Third, advances in vision–language models (VLMs) have demonstrated their potential to enhance access to visual content without human assistance; however, they remain limited in usability and in providing live object descriptions driven by hand–object interactions. 
This motivated our approach of using hands as information cursors to proactively deliver essential visual information beyond the repetitive speech prompts of current AI-enabled ATs. Below, we discuss insights from these domains that shaped the design of {\name}.

\subsection{Hand Interactions as Intent Cues}\label{RW_HandIntentCues}
Hands provide a natural and intuitive interface for interacting with the physical world, effectively conveying users’ intentions~\cite{bolt1980put}, actions~\cite{ma2016going}, and objects or areas of interest~\cite{cai2018understanding, Zhou2022UIST}. Hand gestures are highly expressive and support a wide range of tasks for the general population, including animation creation and authoring~\cite{MagicalHands, Saito2021, PoseTween, TimeTunnel}, mode switching~\cite{Surale2019}, typing~\cite{STAR, ATK}, and object manipulation~\cite{GesturAR, kolaric2008direct, lee20083d, yousefi20163d, mendes2014mid, ong2011augmented, Ubii}.
Beyond visual interactions, hands also play a crucial role in nonvisual exploration. 
For BLV individuals, tactile exploration strategies vary widely and include bimanual, unimanual, and alternating approaches~\cite{wijntjes2008influence, TactileFixations}, which demonstrated the adaptability of hand use strategies to different information needs.
However, when considering the social acceptability of hand interactions, on-body gestures performed within the hands, such as tapping or swiping a finger across one's opposite palm, are generally preferred.
Unlike bodily gestures (e.g., making an 'OK' sign, waving)~\cite{costa2019factors}, these gestures are more discreet, socially acceptable, and feel natural in everyday contexts, such as quickly checking for new messages while commuting~\cite{Oh2014, Oh2017}. 
Drawing from these works, in {\name}, we also considered unique and usable hand interactions for accessing information.

\subsection{Current Use of Hand Interactions for Assistive Technologies}\label{RW_HandInteractionsForAT}
Hand-based interactions have been explored in both commercial ATs and prior research. For instance, BLV individuals commonly access digital information through touch gestures on smartphones. Swipe gestures, for example, enable screen navigation, such as swiping left or right for word-by-word reading, or using a two-finger swipe up in screen readers like TalkBack~\cite{TalkBack} or VoiceOver~\cite{VoiceOver} to read from the top of the screen.
While these methods are effective in digital contexts, comparable approaches for accessing information of physical objects remain limited, often requiring photo capture followed by a question–answering process.
Though tactile exploration can support object understanding~\cite{goldreich2003tactile, TactileFixations}, many rich visual features, such as labeled texts, colors, or intricate patterns, remain inaccessible through touch alone. 

To bridge this gap, prior research has explored using the hands and fingers as information cursors to access visual information in real-time~\cite{Guo2018Cursor}. For instance, prior systems, such as VizLens~\cite{VizLens}, StateLens~\cite{StateLens}, and FetchAid~\cite{FetchAid}, support interactions with appliance control panels by allowing users to point to interface elements that are subsequently read aloud.
Finger-mounted camera systems have been explored as a means of supporting BLV users in accessing visual details, including text~\cite{Findlater2015, Stearns2017, stearns2015design, FingerReader2}, currency and barcodes~\cite{EyeRing}, and clothing color and texture, while maintaining tactile feedback for hands-on exploration~\cite{Medeiros2017, FingerClothing}.
Building on this direction, Lee et al. \cite{lee2021leveraging, lee2020hand, HandsHoldingClues, Lee2019ASSETS} proposed custom models that leverage hand position to localize objects of interest for more effective intent disambiguation and camera alignment. 

Despite these advances, existing systems (Table~\ref{tab:cursors}) often rely on a limited set of gestures and hand-held devices for photo capturing, and provide only single or limited forms of visual feedback (e.g., text, color).
In contrast, enabled by an integrated hand recognition and description generation pipeline, {\name} offers a fluid, hands-free, and integrated experience by delivering live, rich object descriptions driven by hand-object interactions. 
For instance, BLV users can hold or touch an object with one hand to obtain rich visual details, point the object with another to read colors, or perform a swipe-up gesture to access its available texts. 
Such natural and expressive information access was lacking in prior systems.

\subsection{Visual Descriptions with VLMs}
Beyond relying on remote sighted assistance~\cite{bemyeyes, aira} or crowdsourcing~\cite{tseng2022vizwiz, vizwiz}, where human agents may not always be available, recent advancements in VLMs have enabled applications that allow BLV individuals to easily submit image-description queries to AI-powered VQA systems~\cite{chatGPT, bemyai, envisionai, imageexplorer, XuImageExplorer} or receive real-time visual descriptions from live video AI systems~\cite{worldscribe, chatGPTVideo}.
These technologies promote the independence and autonomy of BLV individuals without requiring sighted assistance. We discuss them below.

\subsubsection{Image Capture and Visual Question Answering}\label{RW_VQA}
Current AI-powered visual description systems require users to capture photos and engage in dialogue with AI assistants to obtain specific visual details~\cite{chatGPT, bemyai, envisionai, imageexplorer, XuImageExplorer}.
This process of photo capturing and turn-taking can be laborious and time-consuming. 
For example, taking pictures demands precise camera alignment to ensure the object of interest is within the frame~\cite{vizwiz, Adams2013, Marynel2012, Jayant2011, Kacorri2017ASSETS}, often involving repeated trial and error.
Although cameras with a wider FoV may help mitigate this issue~\cite{VisPhoto}, users must still explicitly specify their needs and interact with the AI to obtain desired details.
This turn-taking VQA process is further challenged by the dynamic nature of the real world, where generated descriptions can quickly become outdated as the environment changes.

\subsubsection{Live Video Feed and Generative Descriptions}\label{RW_liveDescriptions}
Building on photo-taking, ChatGPT’s Advanced Voice with Video~\cite{chatGPTVideo} enables a conversational approach to retrieving visual information through a live video feed. However, instead of proactively delivering essential details, it depends on continuous speech prompts from the user~\cite{assets25}, which may introduce turn-taking delays, increase effort, and raise concerns about privacy and social acceptability. To overcome this lack of proactivity, WorldScribe~\cite{worldscribe} provides live visual descriptions that dynamically adapt to camera motion and the captured visual content.
For example, WorldScribe~\cite{worldscribe} enhanced users’ environmental awareness by providing brief object labels as the camera panned across the surroundings, and offered richer visual details when the camera focused on a specific scene.
In contrast to environmental understanding, our work explores using \emph{hand gestures} as information cursors to proactively describe objects based on how the user is interacting with them, enabling more responsive, intuitive, and fine-grained \emph{object understanding} in real time. 

\section{{\name}}
{\name} is a system that provides live, rich object descriptions based on the user's hand interactions with physical objects.
It detects three types of hand gestures and identifies hand activities in each frame (\emph{Hand Gesture Recognition Layer in Section~\ref{hand_gesture_recog_layer}}). 
Then, {\name} extracts keyframes from live video stream based on these hand activities (\emph{Keyframe Extraction Layer} in Section~\ref{keyframe_layer}) to generate multiple forms of feedback, including hand states, object color, available texts, and brief and detailed object descriptions, and object comparisons (\emph{Description Generation Layer} in Section~\ref{description_generation_layer}). We describe our design goals and implementation details below.

\begin{figure}[h]
\begin{center}
\includegraphics[width=\linewidth, alt={The figure illustrates different classes of hand–object interactions that TouchScribe recognizes. Panel (a) shows continuous states, including holding, touching, pointing, and when the object is out of view, as well as discrete gestures such as hold+point and hold+swipe-up. Panel (b) depicts additional interaction events, such as gesture changes, flipping the object, or user-initiated queries. Panel (c) shows the range of descriptions provided by the system, including hand states, object labels, visual details, color labels, text labels, and visual comparisons.}]{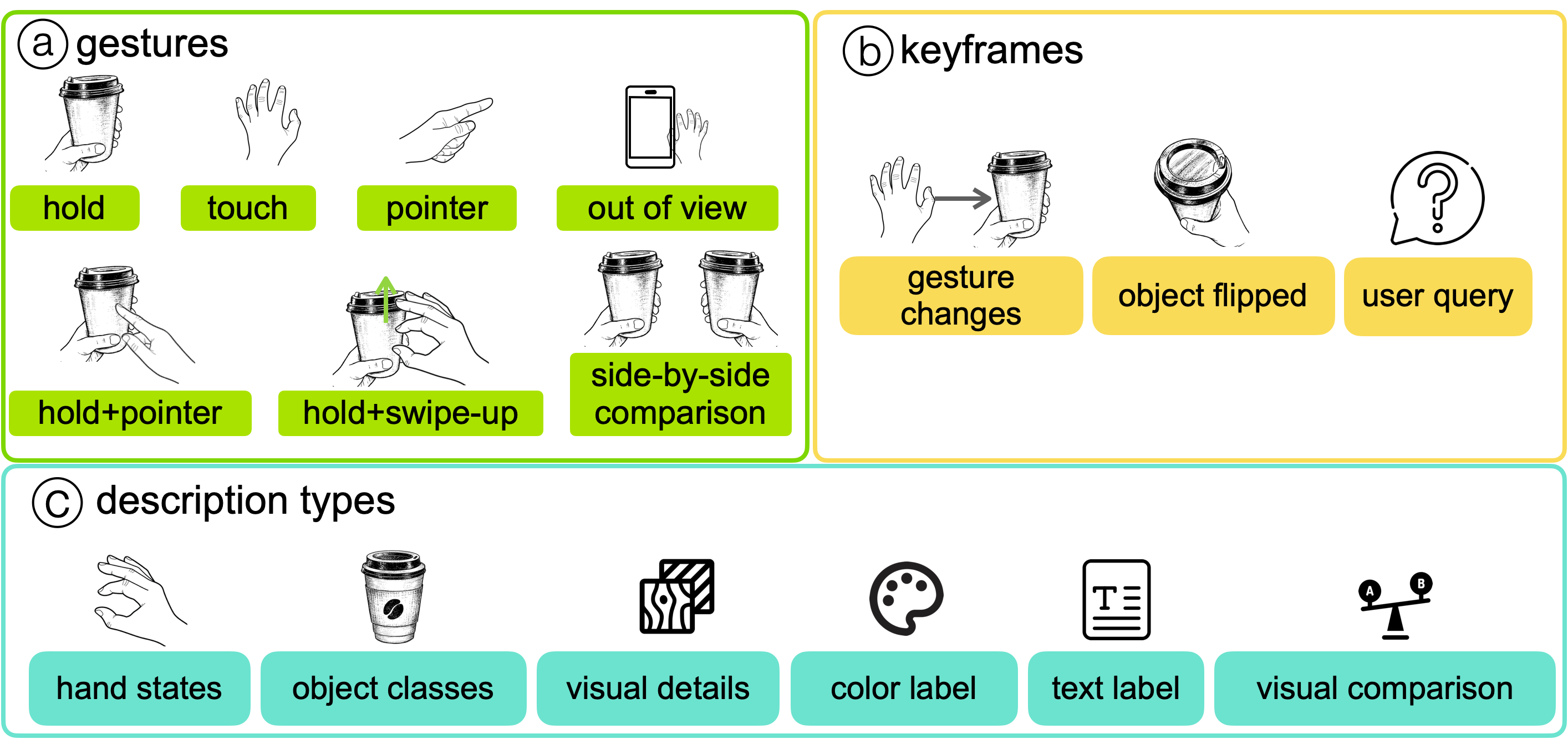}
\vspace{-1.7pc}
\caption{Overview of the variety of gestures, timings to extract keyframes, and description types supported by {\name}.
}
\vspace{-1.7pc}
\label{fig:classes}
\Description{The figure illustrates different classes of hand–object interactions that TouchScribe recognizes. Panel (a) shows continuous states, including holding, touching, pointing, and when the object is out of view, as well as discrete gestures such as hold+point and hold+swipe-up. Panel (b) depicts additional interaction events, such as gesture changes, flipping the object, or user-initiated queries. Panel (c) shows the range of descriptions provided by the system, including hand states, object labels, visual details, color labels, text labels, and visual comparisons.}
\end{center}
\end{figure}

\subsection{Design Goals}
{\name} is designed based on three primary goals inspired by prior work:

\textbf{G1 - Supporting common and usable gestures.} 
Hand interactions serve as valuable intent cues for disambiguating objects of interest~\cite{lee2021leveraging, lee2020hand, Lee2019ASSETS, HandsHoldingClues} and indicating the locations of relevant information (e.g., text~\cite{VizLens, StateLens, FetchAid} or color~\cite{Medeiros2017, FingerClothing}). However, because hand–object interactions vary across individuals and contexts~\cite{TactileFixations}, as an initial step, {\name} should demonstrate a set of common and usable gesture types for BLV individuals.

\textbf{G2 - Supporting proactive and real-time feedback.} 
Given the current strengths and limitations of photo-capturing and VQA-based approaches~\cite{Xie2025CHI, assets25}, which provide access to specific information but introduce delays and turn-taking overhead, {\name} should primarily emphasize proactive feedback while still supporting VQA when needed. Moreover, {\name}’s descriptions should be closely synchronized with hand interactions, minimizing latency between touch and audio output to enhance the overall user experience.

\textbf{G3 - Conveying system-perceived states of hand-object interactions.}
Given that camera aiming has long posed challenges for BLV users~\cite{vizwiz, Adams2013, Marynel2012, Jayant2011}, it is essential to clearly communicate whether the system has detected users’ hands and what it has recognized, enabling users to take appropriate follow-up actions.

\subsection{Gestures to Access Visual Information}\label{gesture_set}
To fulfill \textbf{G1}, {\name} supports six gestures (Figure~\ref{fig:classes}a), categorized along two dimensions: \emph{\textbf{(i) familiarity}}, gestures that are \emph{common} versus those \emph{unique} to BLV users, and \emph{\textbf{(ii) gesture nature}}, gestures that are \emph{continuous} versus \emph{discrete}.
These gestures are informed by prior research on commonly used assistive technologies or discreet on-body gestures~\cite{Oh2014, Oh2017}. 
Each gesture maps to a distinct prompt for VLMs to generate corresponding descriptions (Details are in Section~\ref{description_generation_layer}).

\begin{enumerate}[leftmargin=0.5cm]
    \item \textbf{Hold an object with a single hand \emph{(common \& continuous)}} – Holding an object of interest is a common practice for examining its visual or tactile details, such as reading nutritional information on a bottle or exploring its shape.
    
    \item \textbf{Touch an object with a single hand \emph{(common \& continuous)}} – Touching an object with a few fingers is common for indicating an object of interest in sighted interaction~\cite{Zhou2022UIST}, and is also widely used in tactile exploration by BLV people~\cite{TactileFixations}.

    \item \textbf{Hold or touch explore an object with both hands \emph{(unique \& continuous)}} – Using both hands to explore objects through touch is a common tactile exploration pattern among BLV individuals, particularly when interacting with flat or textured surfaces such as tactile graphics~\cite{goldreich2003tactile, TactileFixations}.

    \item \textbf{Hold or touch objects side-by-side with both hands \emph{(common \& discrete)}} – When comparing similar items, such as ingredient labels on two bottles or subtle shape differences between boxes, people often place or hold them side by side to facilitate comparison.

    \item \textbf{Hold an object in one hand and point with another hand to reveal visual details \emph{(unique \& discrete)}} – Pointing gestures are common for BLV people to access specific information, such as color~\cite{FingerClothing}, text~\cite{VizLens, Findlater2015, stearns2015design, Stearns2016, Stearns2017, FingerReader2, StateLens}, or texture~\cite{MagicFinger, FingerClothing}. 

    \item \textbf{Hold an object in one hand and two-finger swipe up with another hand to read texts. \emph{(unique \& discrete)}}
    Two-finger swipe-up gestures are commonly used in screen readers such as iOS VoiceOver~\cite{VoiceOver} and Android TalkBack~\cite{TalkBack} to read on-screen text from top to bottom. Because the exact locations of text are often unknown to BLV people, we adapt this gesture to enable access to available text on an object.
\end{enumerate}

\begin{figure*}[t]
\begin{center}
\vspace{-1pc}
\includegraphics[width=\linewidth, alt={The diagram is divided into three main stages, labeled (a), (b), and (c), each enclosed by a dashed blue box. Information flows from left to right, starting with a smartphone camera feed and ending with textual descriptions of an object. In stage (a), gesture recognition for live streams is performed. On the far left, a smartphone streams frames that are passed into the pipeline. Each frame undergoes gesture recognition, where hand landmarks are extracted and classified into predefined categories such as “left: hold” and “right: out of view.” A temporal smoothing module then aggregates multiple frames to filter out noise and identify stable gesture states, producing a keyframe. The example shows the stable state “left: grab” and “right: out of view.” In stage (b), object extraction is applied to the selected keyframe. The Hands23 model analyzes the image to infer whether a hand is in contact with an object. The result indicates “left: contact” and “right: None.” The contacted object is then cropped from the frame, producing a clear image of the object, which in this case is a spice bottle. In stage (c), description generation is performed by multiple vision-language models (VLMs). The cropped object image and contact information are passed to different models running in parallel. Moondream generates an action-based description, for example: “Your left hand is holding a spice bottle.” GPT-4o produces a visual recognition description such as “The bottle is chili lime seasoning…,” and a text-focused version of GPT-4o extracts the printed label, producing “Texts: Trader Joe’s Chili Lime Seasoning Blend…” This stage demonstrates how combining different models yields rich descriptions that include both physical interactions and object details.}]{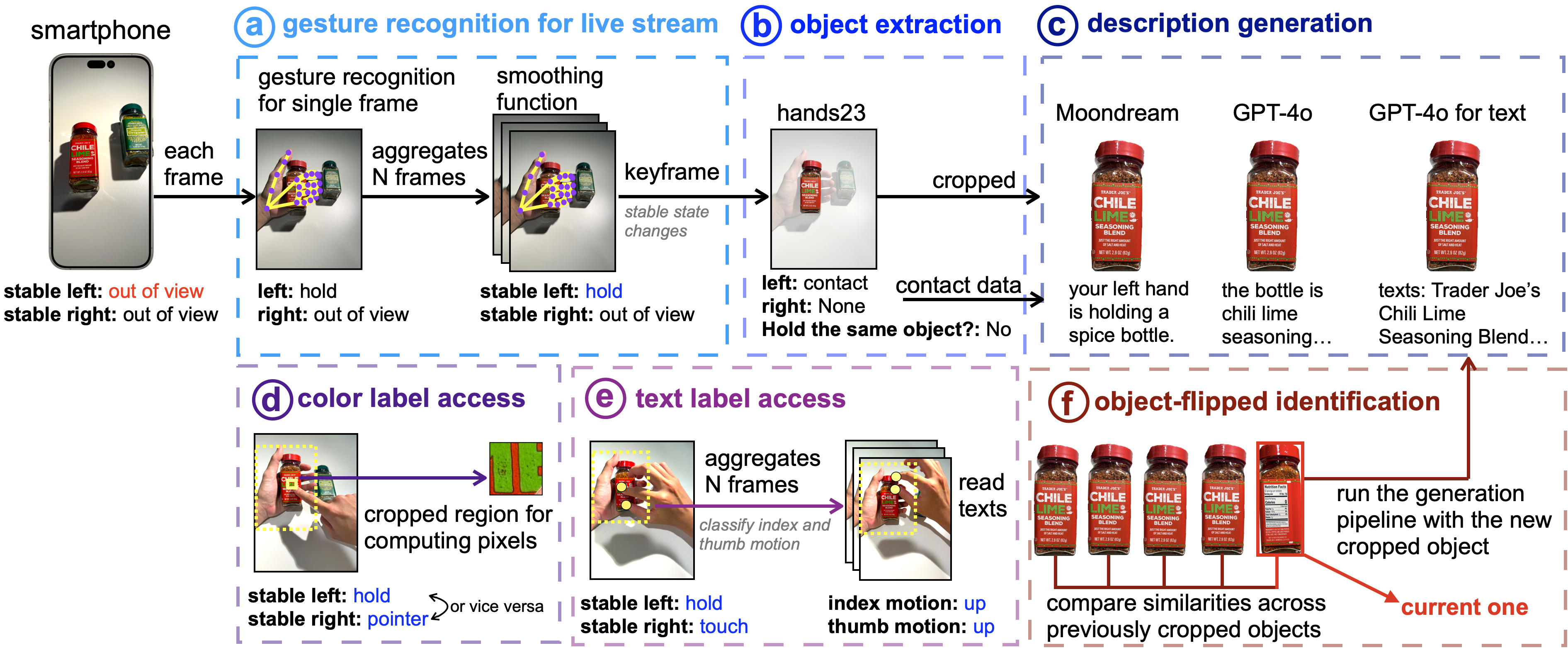}
\vspace{-1.5pc}
\caption{{\name} System Diagram.
(a) {\name} performs gesture recognition on live video streams. For each camera frame, hand landmarks are extracted with Google MediaPipe~\cite{mediapipe} and classified into predefined gesture categories. A temporal smoothing module then aggregates multiple frames to produce stable keyframes and gesture states.
(b) For each keyframe, Hands23~\cite{hand23} infers object contact. The contact data, together with a cropped image of the object, is passed to VLMs for further processing.
(c) VLMs, including Moondream~\cite{moondream} and GPT-4o~\cite{gpt4o}, are executed in parallel to generate rich object descriptions.
(d) When one stable state is \emph{hold} and the other is \emph{point}, {\name} reads the color of the small region the finger is pointing to.
(e) When one stable state is \emph{hold} and the other is \emph{touch}, {\name} tracks finger motion and reads the text once both fingers move up.
(f) {\name} also maintains a history of cropped objects and identifies flipped instances by comparing image similarity, and re-runs the generation pipeline on the updated crop.
}
\vspace{-1pc}
\label{fig:system_diagram}
\Description{The diagram is divided into three main stages, labeled (a), (b), and (c), each enclosed by a dashed blue box. Information flows from left to right, starting with a smartphone camera feed and ending with textual descriptions of an object. In stage (a), gesture recognition for live streams is performed. On the far left, a smartphone streams frames that are passed into the pipeline. Each frame undergoes gesture recognition, where hand landmarks are extracted and classified into predefined categories such as “left: hold” and “right: out of view.” A temporal smoothing module then aggregates multiple frames to filter out noise and identify stable gesture states, producing a keyframe. The example shows the stable state “left: grab” and “right: out of view.” In stage (b), object extraction is applied to the selected keyframe. The Hands23 model analyzes the image to infer whether a hand is in contact with an object. The result indicates “left: contact” and “right: None.” The contacted object is then cropped from the frame, producing a clear image of the object, which in this case is a spice bottle. In stage (c), description generation is performed by multiple vision-language models (VLMs). The cropped object image and contact information are passed to different models running in parallel. Moondream generates an action-based description, for example: “Your left hand is holding a spice bottle.” GPT-4o produces a visual recognition description such as “The bottle is chili lime seasoning…,” and a text-focused version of GPT-4o extracts the printed label, producing “Texts: Trader Joe’s Chili Lime Seasoning Blend…” This stage demonstrates how combining different models yields rich descriptions that include both physical interactions and object details.}
\end{center}
\end{figure*}

\subsection{Implementation Details}\label{implementation_details}
To enable {\name} to provide real-time feedback (\textbf{G2}), we trade off different factors to maximize the computing speed while maintaining decent accuracy (See Section~\ref{technical_evaluation}), such as the choices of the models, or the frame size. 
{\name} servers include a local server running on a MacBook M4 Max and a remote server with two embedded Nvidia GeForce RTX 4090 GPUs. 
{\name} uses a neck mount with an attached iPhone 13 Pro (Figure~\ref{fig:setup}). The smartphone offers more APIs than emerging smart glasses at the time of development, and greater flexibility in selecting frame resolution for real-time use and camera FoV for coverage.
The TouchScribe iOS app uses the wide lens, the 13 mm-equivalent rear camera with an approximately 120° FoV, whereas the standard wide lens (26 mm-equivalent) offers a 77° FoV. 
It streamed the video frames (width: 720, height: 960, configured to retain approximately 70\% image quality) to the local server through a Socket connection.  
Google MediaPipe~\cite{mediapipe}, the hand gesture recognition model and finger motion classification model, runs on the local server and achieves around 6 frames per second (FPS). 
Other models that require higher computational resources run on the remote server, including Hands23~\cite{hand23} for detecting hand–object contacts ($F_1$-score=79.1), SigLIP~\cite{zhai2023sigmoid} for generating image embeddings, and Moondream~\cite{moondream} for producing brief object descriptions.

\subsection{Hand Gesture Recognition Layer}\label{hand_gesture_recog_layer}
In this layer, {\name} aims to recognize the aforementioned hand gestures in a lightweight manner to support real-time performance alongside other models for live visual descriptions (\textbf{G2}).
To achieve this, we fine-tune a hand gesture classification model and a finger motion classification model, which identify gestures and finger movements based on hand landmarks detected using Google MediaPipe~\cite{mediapipe}. We describe these models in detail below.

\subsubsection{Hand gesture classification model}\label{hand_class_model}
We fine-tuned a publicly available keypoint classification model \cite{hand_detection_model} to adapt its model structure for our supported gesture set. 
The model takes a 2D keypoint vector as input and outputs a hand gesture class. 
Specifically, the input vector has a dimensionality of 42 corresponding to the $(x, y)$ coordinates of 21 hand keypoints extracted from Google MediaPipe~\cite{mediapipe}. 
The output includes three gesture categories, \emph{touch}, \emph{hold}, and \emph{point}, for each hand. 
Additionally, a gesture is labeled as \emph{out of view} when no hand keypoints are detected by Google MediaPipe~\cite{mediapipe}.
This results in a total of four classes for each hand.

\subsubsection{Finger motion classification model}
To support the two-finger swipe-up gesture described in Section~\ref{gesture_set}, we fine-tuned a finger motion classification model from the same publicly available repository~\cite{hand_detection_model}.
The model takes as input a time-series history of a fingertip's 2D coordinates $(x, y)$, sampled every 16 frames and resulting in a flattened input vector of size 32. 
The output includes three finger motion gesture categories, including \emph{static}, \emph{up}, and \emph{down}.
This model is executed only when one hand is in the \emph{hold} state and the other is in the \emph{touch} state, with both the index finger and thumb of the \emph{touch} hand located within the bounding box of the \emph{hold} hand (Figure~\ref{fig:system_diagram}e).

\subsection{Keyframe and Object Extraction Layer}\label{keyframe_layer}
In this layer, {\name} identifies keyframes when users perform new gestures or flip an object, signaling the need for updated descriptions (\textbf{G3}). 
A key challenge arises from the intermittent predictions produced by the gesture recognition models due to real-time performance requirements (\textbf{G2}), which may reduce accuracy and lead to false positives.
To mitigate this issue, {\name} applies a temporal smoothing function that analyzes consecutive frames to infer a stable gesture state for each hand.

First, {\name} verifies whether the past $x$ gestures of either hand consist of a single gesture class repeated at least $t$ times (e.g., \emph{hold}, \emph{touch}, \emph{point}, or \emph{out of view}). If this condition is satisfied, that gesture is assigned as the stable gesture state. Otherwise, {\name} checks whether the previous stable gesture state appears within the last $n$ frames and retains it if so. If neither condition holds, the most frequent gesture in the last $n$ frames is selected as the current gesture. Based on our apparatus and empirical tests, we set $x = 12$, $n = 6$, and $t = 4$. Whenever the stable gesture state transitions to either \emph{hold} or \emph{touch}, the corresponding frame is marked as a keyframe and sent to the Hands23~\cite{hand23} model to identify hand–object contact details, supplementing prompt data for the description generation pipeline (Figure~\ref{fig:system_diagram}b).

In addition, when the stable gesture of either hand remains as \emph{hold} or \emph{touch} across keyframes, {\name} analyzes whether the object is unchanged by periodically cropping the object image and computing the cosine similarity between the current image embedding and those from the previous $s$ samples (Figure~\ref{fig:system_diagram}f). If the similarity scores with all $s$ prior samples fall below a threshold $u$, the frame is marked as a keyframe, which indicates a potential change or flip of the object. 
Based on our apparatus and empirical tests, we set $s = 4$ and $u = 0.85$.

The extracted keyframes and objects are passed to VLMs to generate hierarchical feedback and descriptions. 
The structure of these prompts and outputs is detailed in the following section.

\subsection{Description Generation Layer}\label{description_generation_layer}
In this layer, {\name} generates descriptions with adaptive levels of detail based on the user's hand-object interactions. 
To achieve this, {\name}integrates outputs from multiple components, including gesture states from the hand gesture recognition models, hand–object contact information inferred by the Hands23~\cite{hand23} model, and the extracted keyframes and objects.
{\name} dynamically incorporates these details to construct descriptions or prompts for VLMs to generate rich object details. 
We detail each type of description and its generation process below.

\textbf{Hand-State Feedback.} 
Hand-state feedback helps users assess whether the hands are correctly captured and identified (\textbf{G3}). 
Whenever the user's hands are detected within the camera view, {\name} generates feedback such as \PROMPT{I see your \texttt{\{which\_hand\}} hand} to help users confirm the presence of their hands in the frame, where \PROMPT{\texttt{\{which\_hand\}}} is dynamically assigned as \emph{left}, \emph{right} or \emph{both}. 
Then, {\name} describes the perceived stable gesture states, for example: \PROMPT{Your \texttt{\{which\_hand\}} hand is/are \texttt{\{gesture\}}ing} or \PROMPT{You flipped or changed the object.}, where \PROMPT{\texttt{\{gesture\}}} is dynamically assigned based on the recognized stable gesture state, including \emph{hold}, \emph{touch} and \emph{point}.
The two feedback are combined when they are temporally close to reduce repetition, such as \PROMPT{I see your right hand is pointing.}

\textbf{Brief Object Descriptions.}
The brief description helps users quickly assess what the object is, whether it is of interest, and whether they want to learn more. 
Given a keyframe, {\name} first applies the Hands23 model~\cite{hand23} to obtain hand–object contact information, including which hand (or both) is in contact and a cropped image of the object (Figure~\ref{fig:system_diagram}b).
When contact is detected, {\name} generates prompts such as \PROMPT{What is my \texttt{\{which\_hand\}} hand touching?} with a cropped object image.
This prompt is then passed to Moondream~\cite{moondream}, a lightweight VLM that produces concise descriptions with low latency.
Example outputs include \PROMPT{Your right hand is touching a bottle of seasoning.} and \PROMPT{Both your hands are touching a laptop.}

\textbf{Detailed Object Descriptions.}  
Detailed descriptions enable users to access fine-grained visual information about objects.
Using the same cropped object image and hand–object contact data provided to Moondream~\cite{moondream}, {\name} also supplies these inputs to GPT-4o~\cite{gpt4o} with a different prompt: \PROMPT{Can you describe the object I am \{gesture\}ing with my \texttt{\{which\_hand\}} hand in detail?} 
This produces descriptions such as \PROMPT{You are holding a white mug decorated with colorful illustrations...}
Although Moondream~\cite{moondream} and GPT-4o~\cite{gpt4o} perform inference in parallel, Moondream generates an initial high-level description first, followed by GPT-4o’s more detailed output due to differences in latency.

\textbf{Available Object Texts.}  
{\name} reads aloud the available text on the object (e.g., expiration date, nutrition facts) once the user performs the \emph{hold+swipe-up} gesture.
Using the same cropped object image, {\name} submits a different prompt to GPT-4o~\cite{gpt4o}: \PROMPT{I am holding the object with my \texttt{\{which\_hand\}} hand. 
Please describe the text line by line. If there is no text, can you just return 'no text on the \{object name\} your \texttt{\{which\_hand\}} hand is \texttt{\{gesture\}}ing.}
We employ GPT-4o~\cite{gpt4o} for its acceptable latency and accuracy of text recognition on low-resolution images compared to other text recognition models.
This approach enables top-to-bottom reading of text on object surfaces, analogous to screen readers such as iOS VoiceOver~\cite{VoiceOver} and Android TalkBack~\cite{TalkBack}. If users trigger the gesture before texts are generated, {\name} responds: \PROMPT{Still processing the text, please try again later.}

\textbf{Comparative Descriptions.}  
This feedback aims to support comparisons between objects with similar tactile features (e.g., shape and size), enabling users to better understand their visual similarities and differences.
When both hands \emph{hold} or \emph{touch} different objects, {\name} crops the corresponding object images using the Hands23 model~\cite{hand23} and prompts GPT-4o~\cite{gpt4o}: \PROMPT{Can you describe the object I am holding with my left hand and the one with my right hand? What are the differences or similarities between them?} This yields outputs, such as \PROMPT{Your left hand holds a red bottle, and your right hand holds a green one. Similarities: Both are Trader Joe’s... Differences: color and texts are different ...} 
Also, {\name} detects when both hands \emph{hold} or \emph{touch} different parts of the same object, supporting users in understanding the object’s spatial layout and visual characteristics (e.g., surface graphics and text), building on prior work~\cite{goldreich2003tactile, TactileFixations}. 
In this case, {\name} prompts GPT-4o~\cite{gpt4o} with full image and instructions: \PROMPT{Can you describe the spatial and visual relationship between the points I am touching, and highlight any visual similarities or differences between them?} Example outputs include \PROMPT{Your hands touch adjacent areas around the bottle, with the left spanning the text... and the right spanning the graphics...}

\textbf{Color Labels.}  
{\name} reports an object’s color when users \emph{hold} it with one hand and \emph{point} to it with the other. Then, {\name} analyzes a small image region near the index fingertip. Based on the fingertip coordinates and hand side (\texttt{left} or \texttt{right}), the system slightly offsets the cropped region (left/up for the right hand and right/up for the left hand) to exclude the finger itself. It then computes the region’s average RGB value and maps it to the nearest named color using the \emph{webcolors} library~\cite{webcolors}.

\textbf{User Query.}  
Lastly, in line with \textbf{G2}, {\name} enables users to invoke a question-answering function via the voice command \PROMPT{Hey <wake word>} and pose queries. {\name} then submits the current video frame along with the user’s question to GPT-4o~\cite{gpt4o} and reads the generated response, similar to existing AI-enabled assistive VQA services such as BeMyAI~\cite{bemyai} and SeeingAI~\cite{seeingai}.

\subsection{Handling Responsiveness of Descriptions to Hand Interactions and Speech Query}\label{responsiveness}
{\name} prioritizes descriptions and interrupts based on different hand–object gestures. For example, invoking the VQA function interrupts any ongoing narration to address the query, after which hand gestures are ignored until the answer is fully delivered. In contrast, discrete gestures for specific visual information, such as \emph{hold+point} for color labels or \emph{hold+swipe-up} for object text, can also interrupt ongoing descriptions.

\begin{table*}[t]
    \centering
    \small
    \caption{Setup and instructions for each scenario. These scenarios differed based on factors such as \textcolor{purple}{Visual Complexity} in object understanding tasks marked as \textcolor{purple}{\emph{Low}}, and \textcolor{purple}{\emph{High}} in purple, and \textcolor{blue}{Information Specificity} in blue (e.g., \textcolor{blue}{\emph{Specific}} vs. \textcolor{blue}{\emph{General}}).}
    \vspace{-1.em}
    \renewcommand{\arraystretch}{1.2} % Adjust row height
    \begin{tabular}{|m{1.5cm}|>{\raggedright\arraybackslash}m{2cm}|>{\raggedright\arraybackslash}m{3.5cm}|>{\raggedright\arraybackslash}m{5cm}|>{\raggedright\arraybackslash}m{1.5cm}|}
        \hline
        \textbf{Image} & \textbf{Scenario} & \textbf{Setup} & \textbf{Instruction to User} & \textbf{Dimensions} \\ \hline
        \raisebox{-0.1\height}{\includegraphics[width=1.5cm, alt={This is a ceramic coffee mug with a smooth, glossy surface and a sturdy handle. Wrapped around the mug is a colorful, cartoon-style illustration celebrating the city of Toronto, Canada. At the top, the word “TORONTO” is written in large, bold, dark blue capital letters. Below and around this, various symbols and landmarks of the city are depicted. There is an orange sun with sharp rays, and next to it a white takeaway coffee cup with a green circle, representing the city’s coffee culture. A tall clock tower with a steep triangular roof stands prominently, symbolizing Old City Hall, and beside it is a curved, modern building resembling Toronto’s current City Hall, with two tall arches spanning across the scene. Bright orange and yellow autumn leaves and trees appear throughout the design, evoking the feel of fall in the city. An orange streetcar travels along tracks, a nod to Toronto’s iconic public transportation. Sports are also featured, including a blue ice skate and a pair of white hockey gloves, showing the city’s connection to ice hockey. Nearby, there’s a blue recycling bin with a maple leaf on it, representing environmental awareness and civic pride. A baseball bat and ball lie in the lower part of the image, referencing the Toronto Blue Jays. Scattered throughout the design are additional orange and blue maple leaves, further emphasizing Canadian identity. The overall style is playful and stylized, using mostly blue, orange, and white tones to create a vibrant tribute to Toronto’s culture, sports, and landmarks.}]{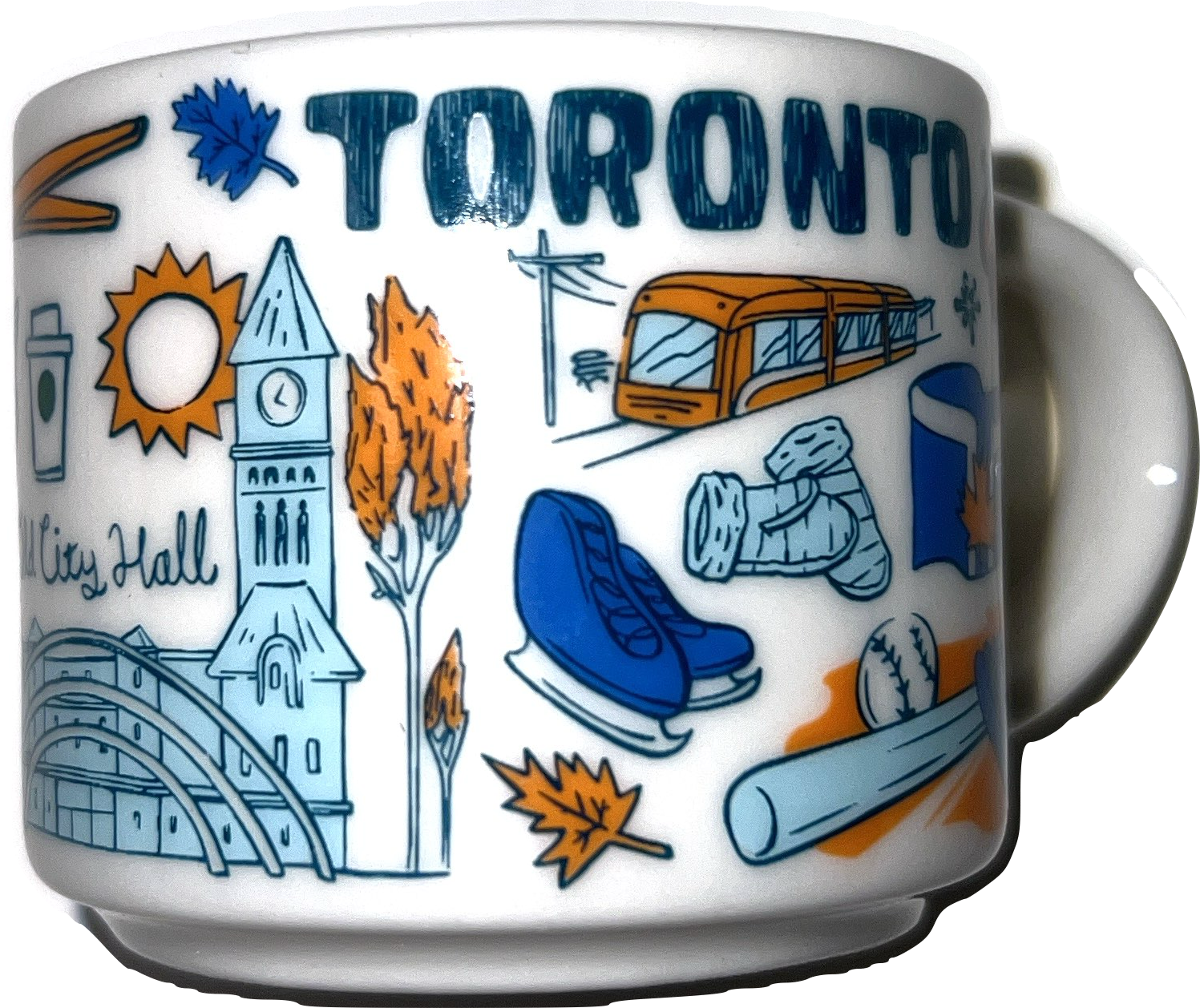}\Description{This is a ceramic coffee mug with a smooth, glossy surface and a sturdy handle. Wrapped around the mug is a colorful, cartoon-style illustration celebrating the city of Toronto, Canada. At the top, the word “TORONTO” is written in large, bold, dark blue capital letters. Below and around this, various symbols and landmarks of the city are depicted. There is an orange sun with sharp rays, and next to it a white takeaway coffee cup with a green circle, representing the city’s coffee culture. A tall clock tower with a steep triangular roof stands prominently, symbolizing Old City Hall, and beside it is a curved, modern building resembling Toronto’s current City Hall, with two tall arches spanning across the scene. Bright orange and yellow autumn leaves and trees appear throughout the design, evoking the feel of fall in the city. An orange streetcar travels along tracks, a nod to Toronto’s iconic public transportation. Sports are also featured, including a blue ice skate and a pair of white hockey gloves, showing the city’s connection to ice hockey. Nearby, there’s a blue recycling bin with a maple leaf on it, representing environmental awareness and civic pride. A baseball bat and ball lie in the lower part of the image, referencing the Toronto Blue Jays. Scattered throughout the design are additional orange and blue maple leaves, further emphasizing Canadian identity. The overall style is playful and stylized, using mostly blue, orange, and white tones to create a vibrant tribute to Toronto’s culture, sports, and landmarks.}} & Understanding an object & Participants were given a cup with colorful graphics and texts. & You got a gift from your friend who just traveled back from a tourist spot. Can you use {\name} to understand this object? In terms of color, texts, and graphics.& \textcolor{blue}{\textbf{General}} \newline \textcolor{purple}{\textbf{Low}} \\ \hline
        \raisebox{-0.1\height}{\includegraphics[width=1.5cm, alt={This image shows two clear glass spice jars from Trader Joe’s placed side by side. Each jar has a plastic screw-on lid and a colorful label on the front. The jar on the left has a red lid and a bright red label with white and green text. It reads “Trader Joe’s Chile Lime Seasoning Blend” in bold letters, with the word “Chile” in white and “Lime” in a vivid green, which is accompanied by small icons of a lime slice and a chili pepper. Beneath that, it says “Just the right amount of salt and heat,” and it contains 2.9 ounces or 82 grams of spice mix. Inside the jar, the seasoning appears reddish-orange with visible flecks of chili and possibly dried lime zest. The jar on the right has a dark green lid and a matching green label with a decorative botanical pattern. It reads “Trader Joe’s Spices of the World Organic Oregano” in white and olive green text. The oregano inside is dry and pale green with small, crumbled leaves visible through the glass. This jar contains 0.45 ounces or 13 grams. Both jars are cleanly designed and stand upright, showing their contents clearly through the transparent glass.}]{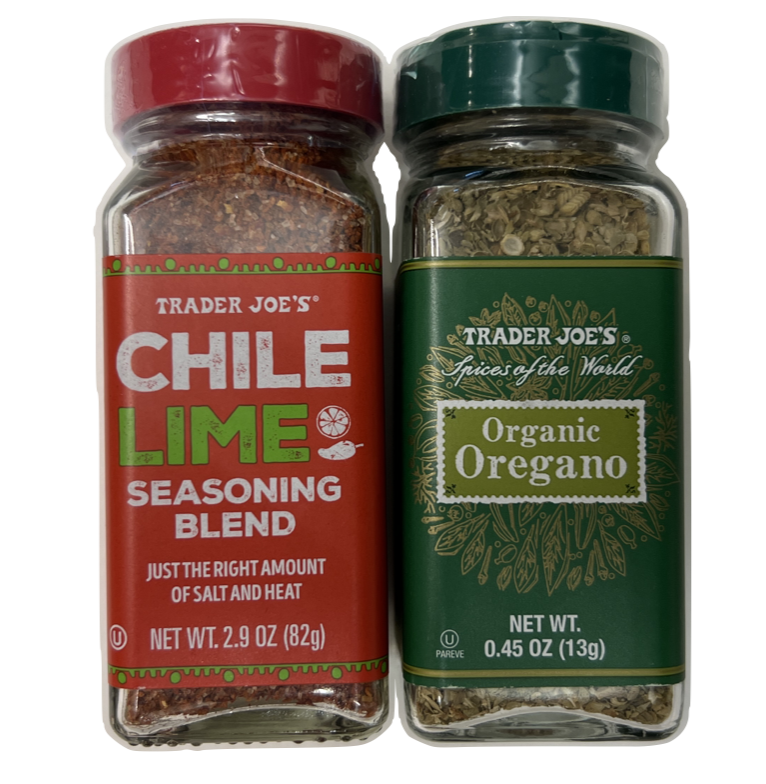}\Description{This image shows two clear glass spice jars from Trader Joe’s placed side by side. Each jar has a plastic screw-on lid and a colorful label on the front. The jar on the left has a red lid and a bright red label with white and green text. It reads “Trader Joe’s Chile Lime Seasoning Blend” in bold letters, with the word “Chile” in white and “Lime” in a vivid green, which is accompanied by small icons of a lime slice and a chili pepper. Beneath that, it says “Just the right amount of salt and heat,” and it contains 2.9 ounces or 82 grams of spice mix. Inside the jar, the seasoning appears reddish-orange with visible flecks of chili and possibly dried lime zest. The jar on the right has a dark green lid and a matching green label with a decorative botanical pattern. It reads “Trader Joe’s Spices of the World Organic Oregano” in white and olive green text. The oregano inside is dry and pale green with small, crumbled leaves visible through the glass. This jar contains 0.45 ounces or 13 grams. Both jars are cleanly designed and stand upright, showing their contents clearly through the transparent glass.}} & Understanding and distinguishing two different spice bottles & Participants were given two spice bottles from Trader Joe's, including one chili lime seasoning with a red label and lid, and another oregano with a green label and lid. & In the grocery store, you have two spice bottles with different labels, colors, and texts. Can you use {\name} to tell the differences and the similarities between them?& \textcolor{blue}{\textbf{General}} \newline \textcolor{purple}{\textbf{High}} \\ \hline
        \raisebox{-0.1\height}{\includegraphics[width=1.5cm, alt={This image shows four small spray bottles of hand sanitizer arranged side by side. The two bottles on the left are from the brand Everyone, labeled in white with red and black text. They are described as ``hand sanitizer spray'' with a ruby grapefruit scent, each containing 2 fl oz (59 mL). The label states that the product is 99.9\% effective against most common germs and is made with plant-based essential oils. The two bottles on the right are from the brand 365 Whole Foods Market, labeled in white with black text and accented in green and purple. These are described as ``Refreshing Spray Hand Sanitizer,'' one scented with cucumber aloe and the other with lavender, each also containing 2 fl oz (59 mL). Both indicate that they kill 99.9\% of germs and include sunflower oil and glycerin to replace moisture. The Everyone bottles are made of darker brown plastic, while the 365 bottles are white, and all feature a spray-top dispenser for easy application.}]{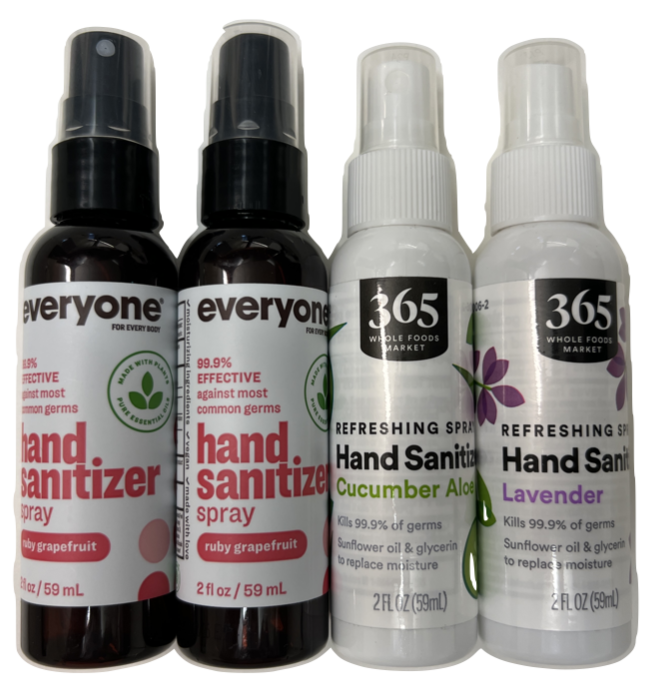}\Description{This image shows four small spray bottles of hand sanitizer arranged side by side. The two bottles on the left are from the brand Everyone, labeled in white with red and black text. They are described as ``hand sanitizer spray'' with a ruby grapefruit scent, each containing 2 fl oz (59 mL). The label states that the product is 99.9\% effective against most common germs and is made with plant-based essential oils. The two bottles on the right are from the brand 365 Whole Foods Market, labeled in white with black text and accented in green and purple. These are described as ``Refreshing Spray Hand Sanitizer,'' one scented with cucumber aloe and the other with lavender, each also containing 2 fl oz (59 mL). Both indicate that they kill 99.9\% of germs and include sunflower oil and glycerin to replace moisture. The Everyone bottles are made of darker brown plastic, while the 365 bottles are white, and all feature a spray-top dispenser for easy application.}} & Understanding and categorizing four spray bottles & Participants were given two identical (from the brand *Everyone*, ruby grapefruit), and the other two were from the same brand (*Whole Foods 365*) but had different scents (cucumber aloe and lavender).& You just got the four spray bottles from a shared storage in your home. Can you use {\name} to categorize them based on their brands and scents?  & \textcolor{blue}{\textbf{Specific}} \newline \textcolor{purple}{\textbf{Low}} \\ \hline 
        \raisebox{-0.1\height}{\includegraphics[width=1.5cm, alt={In the top row, the first juice box on the left is labeled “365 Organic Apple,” with the words “100\% Juice” and “From Concentrate With Added Ingredients.” It comes in a 6.75 fluid ounce size (200 milliliters) and is certified USDA Organic. The packaging is decorated with red and green apples and leaf illustrations. The second juice box is “365 Organic Lemonade,” sweetened with grape juice and also made from concentrate. It has bright yellow slices of lemon and orange on the design, with similar size and USDA Organic certification. The third juice box is another apple juice, labeled “Organic Apple” but with “55\% Less Sugar” and “35 Calories per Box,” using the same volume. The packaging includes similar apple-themed graphics with a blue circular callout noting the reduced sugar. In the bottom row, the three chocolate bars are from the Chocolove brand. Each bar is 3.2 ounces (90 grams) and has a cocoa percentage noted at the top. The first bar is labeled “55\% Cocoa – Almonds & Sea Salt in Dark Chocolate,” with a copper-orange wrapper and images of whole almonds. The second is “65\% Cocoa – Rich Dark Chocolate,” in a deep red wrapper showing a square of glossy chocolate. The third is “70\% Cocoa – Almonds & Sea Salt in Strong Dark Chocolate,” wrapped in black and also showing almonds. Each chocolate bar has a small gold heart emblem, a faux postage stamp graphic, and seals indicating Non-GMO Project Verified and Rainforest Alliance or similar certifications. The bars are styled elegantly, with a handwritten script logo and upscale design elements to evoke a premium feel.}]{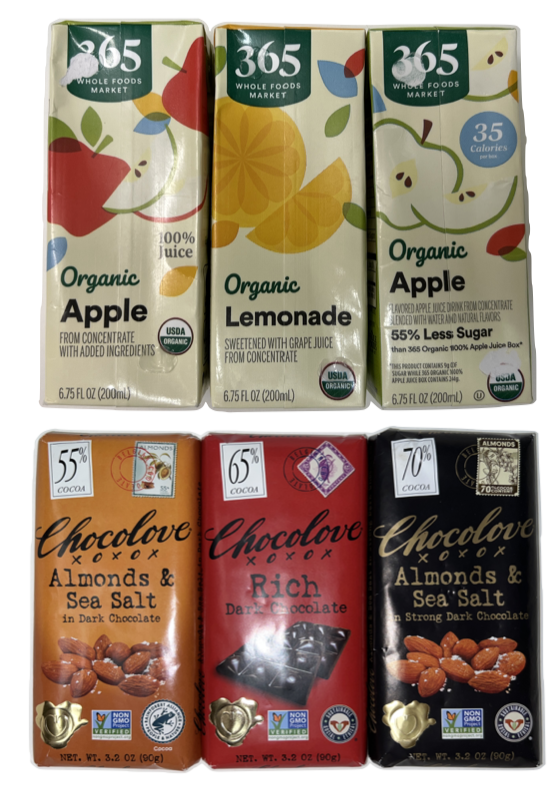}\Description{In the top row, the first juice box on the left is labeled “365 Organic Apple,” with the words “100\% Juice” and “From Concentrate With Added Ingredients.” It comes in a 6.75 fluid ounce size (200 milliliters) and is certified USDA Organic. The packaging is decorated with red and green apples and leaf illustrations. The second juice box is “365 Organic Lemonade,” sweetened with grape juice and also made from concentrate. It has bright yellow slices of lemon and orange on the design, with similar size and USDA Organic certification. The third juice box is another apple juice, labeled “Organic Apple” but with “55\% Less Sugar” and “35 Calories per Box,” using the same volume. The packaging includes similar apple-themed graphics with a blue circular callout noting the reduced sugar. In the bottom row, the three chocolate bars are from the Chocolove brand. Each bar is 3.2 ounces (90 grams) and has a cocoa percentage noted at the top. The first bar is labeled “55\% Cocoa – Almonds & Sea Salt in Dark Chocolate,” with a copper-orange wrapper and images of whole almonds. The second is “65\% Cocoa – Rich Dark Chocolate,” in a deep red wrapper showing a square of glossy chocolate. The third is “70\% Cocoa – Almonds & Sea Salt in Strong Dark Chocolate,” wrapped in black and also showing almonds. Each chocolate bar has a small gold heart emblem, a faux postage stamp graphic, and seals indicating Non-GMO Project Verified and Rainforest Alliance or similar certifications. The bars are styled elegantly, with a handwritten script logo and upscale design elements to evoke a premium feel.}} & Finding products with specific information & Participants were given three carton of juices, including two apple juices (100 \& 35 calories) and one lemonade (100 calories), and three chocolate bars (55, 65, 70\% of cocoa). & You want to find some snacks in a shared pantry, specifically, the chocolate bars with the most cocoa and the apple juice with the fewest calories for your health. Can you use {\name} to help you find them? & \textcolor{blue}{\textbf{Specific}} \newline \textcolor{purple}{\textbf{High}} \\ \hline 
    \end{tabular}
    \Description{}
    \label{tab:image_table}
\end{table*}

\section{Evaluation Methods}
We conducted a user study to qualitatively understand \textbf{How do BLV participants experience and perceive {\name}?} 
We then used the captured videos and interaction data from the study to conduct a technical evaluation for quantitative insights into \textbf{What is the accuracy and latency of {\name}'s descriptions, in response to users' hand-object interactions?}
We detail our methods and results below.

\subsection{Participants}
We recruited eight BLV participants (3 Male and 5 Female) using email lists for local accessibility organizations, prior contacts, and snowball sampling.
Participants aged from 18 to 72 (Avg. 45.5) and described their visual impairment as blind (N=6) or having low vision (N=2).
Most participants had prior experiences using remote sighted assistance and AI-enabled services, such as Orcam~\cite{Orcam}, BeMyEyes~\cite{bemyeyes}, BeMyAI~\cite{bemyai}, Aira~\cite{aira}, or SeeingAI~\cite{seeingai} in their daily lives (Table \ref{tab:demographic}).

\begin{figure}[t]
\begin{center}
\includegraphics[width=\linewidth, alt={The image shows a person seated at a table, wearing a neck-mounted phone holder. Attached to the mount is a smartphone positioned in front of their chest, angled downward so its camera faces the table and the person’s hands. The person is holding two small spice jars—one with a green label and one with a red label. The phone screen displays a live camera view of their hands holding the same two jars, along with an app interface that appears to be generating text descriptions of what the camera sees. The table surface is light-colored, and the setting appears to be a casual indoor environment.}]{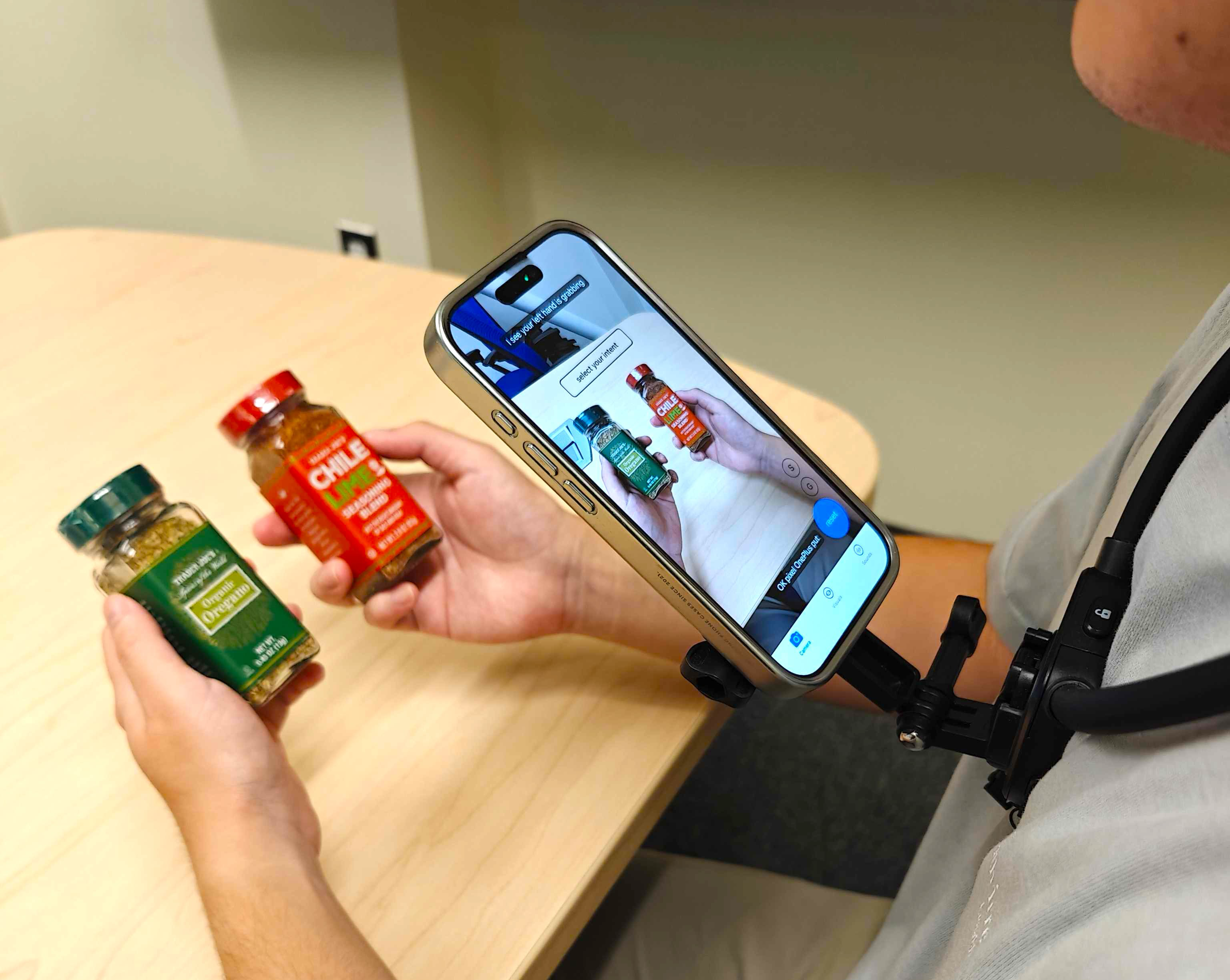}
\vspace{-1.2pc}
\caption{The {\name} prototype setup included an adjustable neck mount with an attached smartphone. During the study, researchers adjusted the mount for each participant to ensure the camera was properly aimed at the table.}
\label{fig:setup}
\Description{The image shows a person seated at a table, wearing a neck-mounted phone holder. Attached to the mount is a smartphone positioned in front of their chest, angled downward so its camera faces the table and the person’s hands. The person is holding two small spice jars—one with a green label and one with a red label. The phone screen displays a live camera view of their hands holding the same two jars, along with an app interface that appears to be generating text descriptions of what the camera sees. The table surface is light-colored, and the setting appears to be a casual indoor environment.}
\vspace{-1pc}
\end{center}
\end{figure}

\subsection{Procedure, Apparatus and Tasks}
The study consisted of two sessions: (i) a \emph{practice session}, designed to familiarize participants with {\name}, and (ii) a \emph{task session}, during which participants completed a series of object understanding and selection tasks. Throughout the study, participants remained seated and interacted with the system using a neck-mounted smartphone (Figure~\ref{fig:setup}).

\textbf{(i) Practice session.} 
Participants were introduced to the hand gestures supported by {\name}, the corresponding feedback and descriptions provided by the system, and the procedure for invoking the VQA function.

\textbf{(ii) Task session.} 
During the task session, participants completed four object understanding and selection tasks with increasing levels of complexity, determined by the number of objects and the specificity of required information~\cite{assets25}. We describe each task below.

\begin{figure*}[t]
\begin{center}
\vspace{-0.5pc}
\includegraphics[width=\linewidth, alt={This chart presents survey results across seven statements. For the statement “I would use this system in my life for these tasks. (usefulness)” responses lean positive with 2 somewhat disagree, 1 neither agree nor disagree, 2 somewhat agree, and 3 strongly agree. For “I felt that the interaction was intuitive. (intuitiveness),” the distribution is 1 somewhat disagree, 1 neither agree nor disagree, 4 agree, and 2 strongly agree. For “I felt in control of how I received information about the object. (agency),” responses include 1 strongly disagree, 1 somewhat disagree, 1 neither agree nor disagree, 2 agree, and 3 strongly agree. For “The system provided a complete description of the object. (coverage),” results are overwhelmingly positive with 1 neither agree nor disagree, 1 somewhat agree, and 6 strongly agree. For “The system provided an accurate description of the object.(accuracy),” the breakdown is 1 somewhat disagree, 2 neither agree nor disagree, 3 agree, and 2 strongly agree. Finally, for “I was able to obtain the specific information I needed. (effectiveness),” responses are 2 neither agree nor disagree, 4 agree, and 2 strongly agree. Overall, the trend across all statements is strongly positive, with agreement and strong agreement dominating.}]{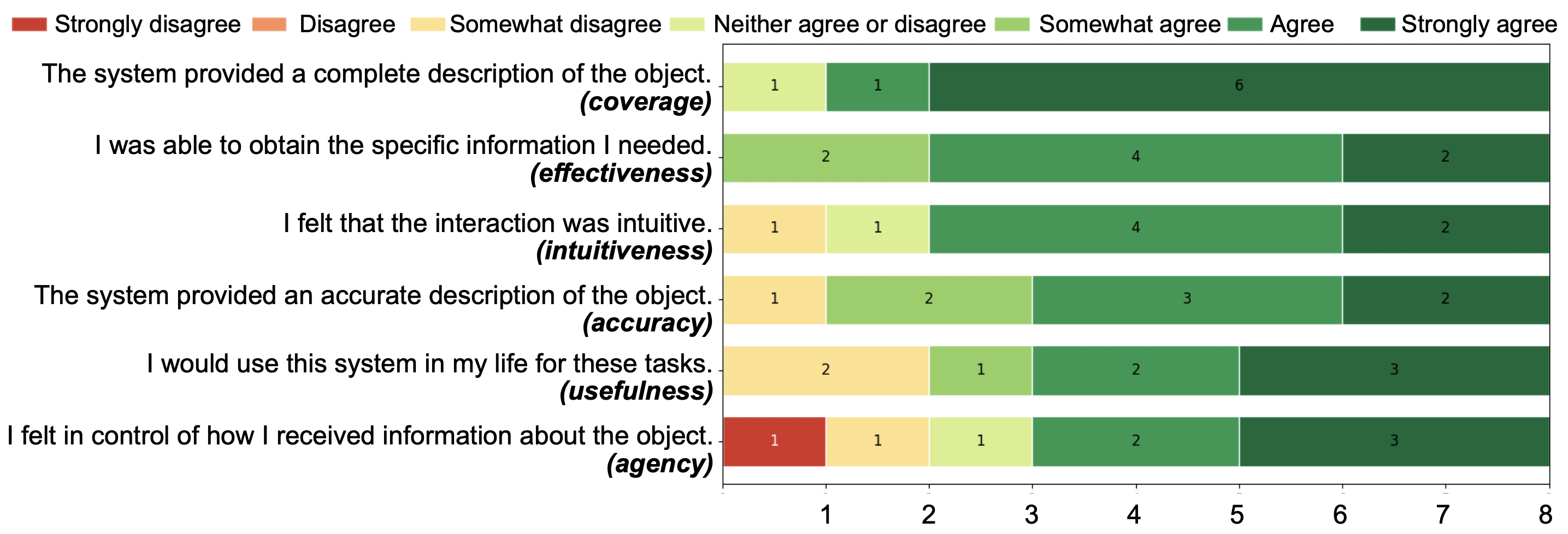}
\vspace{-1.8pc}
\caption{Likert scale questions and aggregated responses of eight participants in our user study. This includes questions about coverage (M=6.5, SD=1.07), effectiveness (M=6, SD=0.76), intuitiveness of gestures (M=5.63, SD1.41), accuracy of descriptions (M=5.5, SD=1.6), usefulness (M=5.5, SD=1.69), and agency of using TouchScribe (M=5.13, SD=2.23).
}
\label{fig:likert_scale}
\Description{This chart presents survey results across seven statements. For the statement “I would use this system in my life for these tasks. (usefulness)” responses lean positive with 2 somewhat disagree, 1 neither agree nor disagree, 2 somewhat agree, and 3 strongly agree. For “I felt that the interaction was intuitive. (intuitiveness),” the distribution is 1 somewhat disagree, 1 neither agree nor disagree, 4 agree, and 2 strongly agree. For “I felt in control of how I received information about the object. (agency),” responses include 1 strongly disagree, 1 somewhat disagree, 1 neither agree nor disagree, 2 agree, and 3 strongly agree. For “The system provided a complete description of the object. (coverage),” results are overwhelmingly positive with 1 neither agree nor disagree, 1 somewhat agree, and 6 strongly agree. For “The system provided an accurate description of the object.(accuracy),” the breakdown is 1 somewhat disagree, 2 neither agree nor disagree, 3 agree, and 2 strongly agree. Finally, for “I was able to obtain the specific information I needed. (effectiveness),” responses are 2 neither agree nor disagree, 4 agree, and 2 strongly agree. Overall, the trend across all statements is strongly positive, with agreement and strong agreement dominating.}
\end{center}
\end{figure*}

\begin{figure}[h]
\begin{center}
\includegraphics[width=\linewidth, alt={This image is a bar chart with six vertical bars, each representing a different category of data. The bars are colored light blue, and each has a black vertical error bar extending above and below, showing variability in the data. Above each bar, the mean (M) and standard deviation (SD) values are displayed. The first bar about "Mental Demand" shows M = 3.19 and SD = 2.45, making it one of the taller bars with moderate variability. The second bar about "Physical Demand" has M = 2.5 and SD = 3.01, indicating a slightly lower mean but higher spread. The third bar about "Temporal Demand" is much shorter, with M = 1.31 and SD = 1.31, suggesting a low mean and smaller variability. The fourth bar about "Performance" is similar in height to the third, with M = 8.56 and SD = 0.9, showing a low mean and the smallest variability of all categories. The fifth bar about "Effort" rises again, with M = 2.5 and SD = 3.08, indicating both a moderate mean and a high spread. Finally, the sixth bar about "Frustration" has M = 2.44 and SD = 3.32, reflecting a moderate mean with the largest variability among the six categories.}]{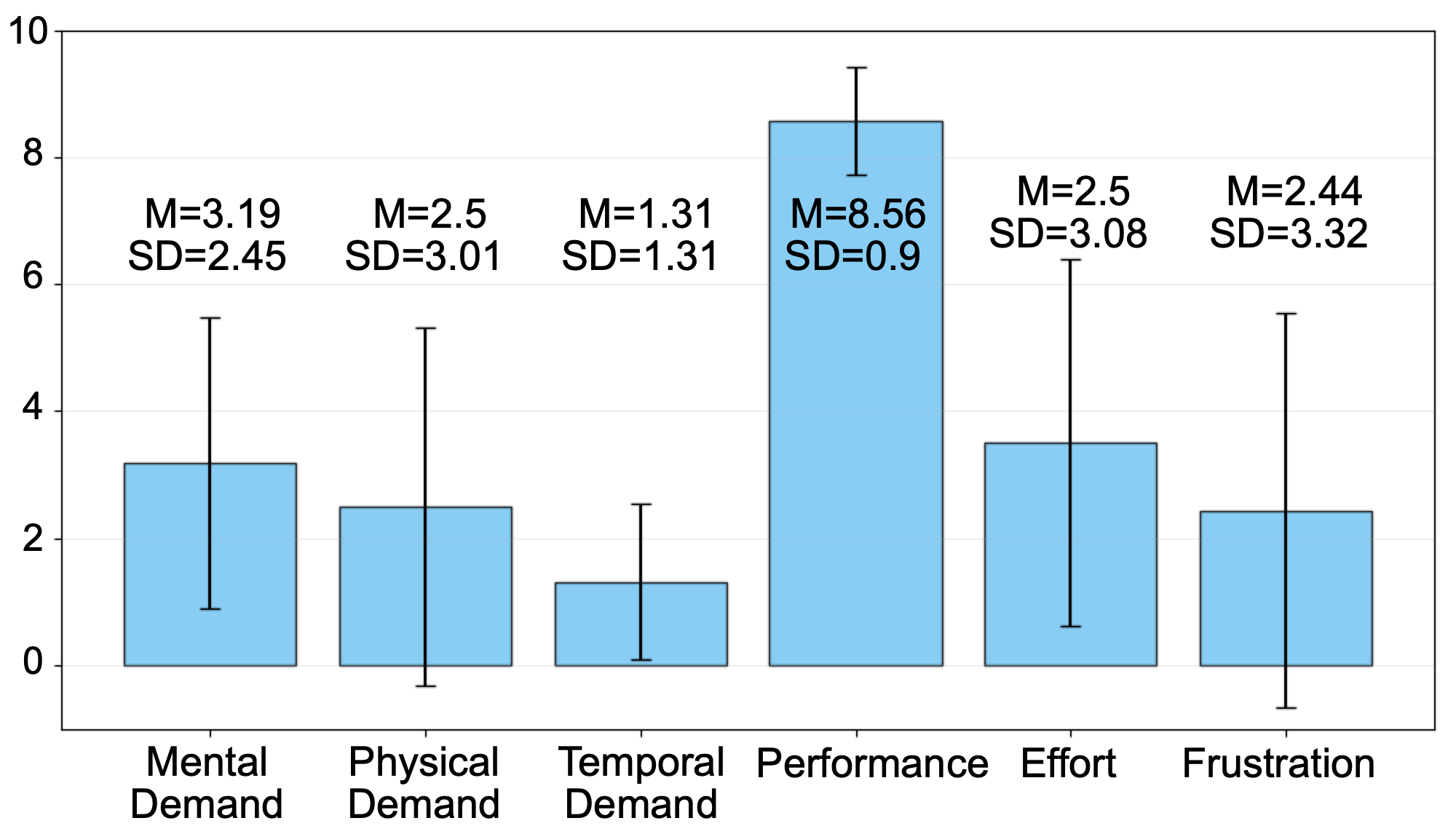}
\vspace{-1.5pc}
\caption{NASA-TLX responses from the user study. Higher scores on the \emph{Performance} dimension indicate better outcomes, whereas lower scores on the remaining dimensions reflect better outcomes.
}
\vspace{-1pc}
\label{fig:nasa}
\Description{This image is a bar chart with six vertical bars, each representing a different category of data. The bars are colored light blue, and each has a black vertical error bar extending above and below, showing variability in the data. Above each bar, the mean (M) and standard deviation (SD) values are displayed. The first bar about "Mental Demand" shows M = 3.19 and SD = 2.45, making it one of the taller bars with moderate variability. The second bar about "Physical Demand" has M = 2.5 and SD = 3.01, indicating a slightly lower mean but higher spread. The third bar about "Temporal Demand" is much shorter, with M = 1.31 and SD = 1.31, suggesting a low mean and smaller variability. The fourth bar about "Performance" is similar in height to the third, with M = 8.56 and SD = 0.9, showing a low mean and the smallest variability of all categories. The fifth bar about "Effort" rises again, with M = 2.5 and SD = 3.08, indicating both a moderate mean and a high spread. Finally, the sixth bar about "Frustration" has M = 2.44 and SD = 3.32, reflecting a moderate mean with the largest variability among the six categories.}
\end{center}
\end{figure}

\begin{enumerate}[leftmargin=1cm]
    \item \textbf{Understanding an object}: Participants were given a cup featuring text and graphics on its surface (Table~\ref{tab:image_table}). The cup was placed on a table, and participants were instructed to use {\name} to obtain descriptions to understand its visual features. The task concluded when participants felt they had sufficiently understood the cup’s visual characteristics and reported their observations to the experimenter.
    
    \item \textbf{Distinguishing two similar objects}: Participants were provided with two seasoning bottles of identical shape but differing in labels, colors, and text. They were asked to identify both similarities and differences between the bottles. The task concluded when participants felt they had sufficiently understood these attributes and reported their observations to the experimenter.
    
    \item \textbf{Sorting four similar objects}: Participants were provided with four bottles of similar shape and size: two identical bottles of the \emph{Everyone} brand (grapefruit scent) and two bottles from the \emph{Whole Foods 365} brand with different scents (cucumber aloe and lavender). They were asked to categorize the bottles by brand and scent. The task concluded when participants indicated they had completed the categorization.

    \item \textbf{Selecting objects based on specified needs}: Participants took part in a shared pantry scenario in which they were asked to locate items based on specific nutritional information. The setup included six products: three chocolate bars with varying cocoa content (55\%, 65\%, and 70\%) and three beverages, two apple juices with 100 and 180 calories, and one lemonade. Participants were instructed to identify the chocolate bar with the highest cocoa content and the apple juice with the fewest calories. The task concluded when participants indicated they had finished.
\end{enumerate}
For each task, objects were randomly placed on the table in front of participants rather than deliberately staged. This allowed the objects to be encountered naturally without excessive search time, as object finding was not the focus of our study.
To support the collection of qualitative insights, participants were encouraged to think aloud and take their time exploring {\name} while completing the tasks.
After completing all tasks, participants responded to a set of Likert-scale questions (Figure~\ref{fig:likert_scale}), completed the NASA-TLX form to assess cognitive load (Figure~\ref{fig:nasa}), and shared their overall experiences.

The entire study lasted about one hour. Participants were compensated for their transportation costs and an additional \$25 for their participation.
This study was approved by the Institutional Review Board (IRB) at our institution.

\subsection{Data Collection and Analysis}
For the user evaluation, we collected participants’ responses to a set of Likert-scale questions across multiple dimensions, including perceived effectiveness, intuitiveness, usefulness, perceived accuracy and coverage of descriptions, and sense of agency when using {\name} (Figure~\ref{fig:likert_scale}). Participants also completed the NASA-TLX questionnaire~\cite{nasatlx} to assess cognitive workload. Additional insights were obtained through open-ended questions in a semi-structured interview, and the entire session was video recorded. Two researchers transcribed the interviews and analyzed the qualitative data using affinity diagramming.

In addition, interactions with {\name} were logged for technical evaluation, including recognized gestures, generated descriptions, and referenced frames (Section~\ref{technical_evaluation}). To analyze these data, we conducted a round-table discussion and annotation session with four members of the research team. The researchers collaboratively reviewed the images and their corresponding descriptions, dividing the workload. Ambiguities or questions raised by any team member were resolved through group discussion.

\subsection{Limitation}
Our study was conducted in a controlled lab environment, with participants seated throughout the sessions due to the study's extended duration. 
We acknowledge that this setting may not fully reflect real-world conditions, where users may interact with cluttered environments or objects that exceed typical hand-grasp ranges. Although the neck-mounted smartphone was designed to approximate an egocentric perspective, it may be impractical for everyday use because of potential social acceptability concerns.
Such issues could be mitigated through alternative form factors, such as smart glasses or more discreet wearable setups (e.g., a yarn lanyard). Additionally, lighting conditions and camera angles were adjusted for each participant to accommodate the limitations of the current hand landmark detection model.
Despite these constraints, our primary goal was to demonstrate the feasibility of delivering live descriptions driven by hand-object interactions. 
We discuss these limitations and potential solutions in Section~\ref{discussion}.

\section{User Evaluation Results}
In general, participants were able to use {\name} to complete a majority of the tasks. 
They commended the accuracy and coverage of the information provided, as well as the intuitive way to access specific details, especially in comparison to the tools they currently use. However, participants also identified several limitations, including latency in retrieving specific information due to the hierarchical feedback design, interruptions triggered by unintentional hand movements, and a learning curve associated with the new interaction techniques. 
We elaborate on these findings below.

\subsection{Overall Task Completion} 
Overall, participants successfully completed the majority of tasks (27 out of 32 tasks), typically within 5-10 minutes, and reported high perceived effectiveness in using {\name} to obtain specific information (M=6.0, SD=0.76).

Specifically, for \emph{Task 1 – Understanding an object}, participants achieved an 87.5\% completion rate by correctly identifying the text and colors on the cup. 
Most used gestures, such as \emph{hold} the cup and flip it around to access surface details, and use and \emph{hold+point} to access its color. 
One exception was P1, who misidentified the interior color as black due to shading while pointing inside the cup.

For \emph{Task 2 – Distinguishing two similar objects}, all participants successfully identified differences in brand names, spice labels, and bottle colors. 
Common strategies we observed included holding both bottles side-by-side for comparative descriptions or examining each bottle individually at a time to verify visual details. 

Similarly, in \emph{Task 3 – Sorting four similar objects}, participants reached a 75\% completion rate. Most participants distinguished the bottles using both color labels and visual descriptions, but some struggled with reading text on curved surfaces, leading to hallucinated or incomplete descriptions. 
This caused confusion for P2 and P6, who did not complete the task.

In \emph{Task 4 – Selecting objects with specified needs}, the completion rate was 75\%. All participants successfully identified the chocolate bar with the highest cocoa content, but some encountered difficulties with the juice selection. For example, P1 was unable to locate the side with the calorie label and gave up, while P3 misremembered the calorie values despite receiving accurate descriptions.

\subsection{Perceived Accuracy, Completeness, and Latency of Descriptions}\label{results_accuracy}
\textbf{Participants found that {\name} provided accurate and comprehensive descriptions; however, the density and prioritization of the information occasionally hindered efficient access.}

Overall, participants perceived descriptions to be accurate (M=5.5, SD=1.6) and complete (M=6.5, SD=1.07), such as \QUOTE{I can get descriptions of bottle, texts on it, and colors too. Without a person or an app like Be My Eyes or Aira, you usually just get one of them and miss the full picture.} (P3) or \QUOTE{It's detailed, descriptive, and reads ingredients verbatim per se} (P2).
P1 also found the coverage of {\name}'s descriptions informative than his current apps: \QUOTE{If I use Seeing AI, I just held it there a minute until it starts reading words. And as soon as I recognized a keyword, I knew what it was. Whereas with [{\name}], it is more. It doesn't just read the 1st word that it comes to, but also recognizes the object like a box of cereal. It's Cheerios, whereas Seeing AI is just gonna start reading randomly, heart health, 100 calories, and great with milk, and then it might say, Cheerios. [{\name}] is recognizing the object, instead of just saying words.}

However, despite the comprehensive coverage of information, participants expressed mixed perceptions regarding the density and prioritization of the spoken content. 
For example, participants noted that hand-state feedback would be more appropriate as \QUOTE{a tutorial at the beginning to understand what it sees (P6),} rather than being presented regularly, which they found somewhat distracting.
Also, P4 felt the transitions between descriptions were smooth, but suggested adding a brief pause in between for easier comprehension: \QUOTE{It was telling me more than what I needed to know at that moment. Maybe consider adding a second or two.}

Furthermore, the hierarchical feedback design, progressing from brief to more detailed object descriptions, presented both advantages and drawbacks. On the one hand, it could slow access to specific details, such as retrieving nutritional information in Task 4, where a direct VQA might be more efficient. 
On the other hand, it helped contextualize information and maintain coherence across descriptions. For example, P3 found the hierarchical structure helpful for distinguishing between similar objects, noting in Task 3: \QUOTE{It said that all of them were hand sanitizers and then went down into more specific information, like this is orange blossom, and this is cucumber. It drills down to the more specific information, and it would be easy to tell which is which.}

\subsection{Perceived Agency, Gesture Intuitiveness, and Hand Constraints}
\label{results_intuitiveness}
\textbf{Participants generally found the gestures intuitive and felt in control when accessing information, though they also noted usability challenges related to hand movements.}

Participants rated the gestures as intuitive (M=5.63, SD=1.41) and reported a sense of control when using {\name} (M=5.13, SD=2.23).
P2, who regularly used OrCam~\cite{Orcam} for text reading, noted that hand-based interaction provided greater control: \QUOTE{The way you move your hand tells the system everything it needs to describe the object. For glasses, you have to chin down, use your nose, and go down towards the text to get everything in the block. This ({\name}) did me a replacement by just holding the object.}

Similarly, P4 appreciated the immediacy of {\name} compared to applications used in daily life, such as Be My AI~\cite{bemyai}: \QUOTE{I just want immediate responses, because Be My AI will take a picture and tell you some basic things. And you need to go to chat for more information. [{\name}] is more immediate. You don't have to go through a chat to do it. That's just right there at the fingertips. Immediate. This would be a good app for people who do not have the patience to mess with chat.}

Participants also found {\name}’s feedback on text detection and object flipping helpful for identifying items whose visual information is distributed across multiple surfaces and not fully visible from a single viewpoint.
This feature reminded participants of grocery shopping experiences in which they needed to locate product barcodes to access digital information using existing assistive applications (e.g., Seeing AI~\cite{seeingai}), suggesting that {\name} could provide practical benefits in such contexts.
As noted by P3, \QUOTE{That's difficult if you don't know where the barcode is located. You need to turn the item in all kinds of ways to get the system to recognize that barcode. With this ({\name}), you don't need to wait for locating the barcode. It just told what this is, and if there is text. So I knew to turn it to the other side.}

During the study, {\name} occasionally misinterpreted idle hand movements or noise in posture detection as intentional input, resulting in false positives and unintended interruptions (see Section~\ref{techEval_hand_gesture}).
Participants noticed these disruptions but generally viewed the system’s sensitivity as a trade-off. 
As P7 said: \QUOTE{It restarted whenever I was even just moving a little bit... but checks and balances...because previous descriptions might go on for too long if it didn't restart.} 
To adapt, some participants intentionally moved their hands out of the camera’s view to pause the system and then brought them back to reset the hierarchical feedback.
We further discuss these limitations and propose future directions for improving gesture recognition and intent disambiguation in real-time settings in Section~\ref{discussion_sensor}.

\subsection{Perceived Cognitive Load and Learning Curve}\label{results_cognitive_load}
\textbf{Participants generally found {\name} usable and easy to learn, though they reported moderate cognitive effort and a noticeable learning curve related to gesture use and hand positioning.}

Although participants rated {\name} as useful (M=5.5, SD=1.69), the task design imposed noticeable but moderate cognitive demands (NASA-TLX: M=3.19, SD=2.45 out of 10), as participants needed to remember descriptions and associate them with the corresponding objects.
As P3 noted, \QUOTE{I had to pay attention to try to remember what it was saying,} and P6 described the experience as \QUOTE{like a memory test.}
Additionally, the walk-up-and-use study design introduced extra effort in learning the mappings between hand gestures and the description categories supported by {\name}.

On the other hand, participants generally perceived the gestures as common and easy to learn; however, the \emph{hold+swipe-up} gesture for accessing text was considered less intuitive. As P1 noted, \QUOTE{I would not guess this unless you told me the inspiration was that (from VoiceOver)}.
In addition, participants reported that positioning their hands within the camera frame required effort. P6 explained, \QUOTE{In theory, it ({\name}) is very quick to learn, which only took us 2 minutes to go through all. In practice, it's a learning process of getting the hand placement just right, because it's a little bit finicky.} 

The combined effects of these challenges, including interruptions from gesture recognition errors and delays introduced by hierarchical feedback, occasionally increased cognitive effort.
However, compared to current practices, participants still perceived hand-based information access as convenient. As P1 remarked, \QUOTE{If you are at the store and you have to continually find ways to read different products, using hands would be easier and more convenient.}
We discuss potential improvements in gesture customization and camera aiming to reduce these demands and enhance {\name}’s usability in Sections~\ref{discussion_customization} and~\ref{discussion_camera}.

\section{Technical Evaluation Results}\label{technical_evaluation}
Using data from the user study, we conducted a technical evaluation of (i) the hand gesture recognition performance of our pipeline in live video stream, (ii) the accuracy of the system-generated descriptions (Table~\ref{tab:accuracy_results}), and (iii) the latency between gesture input and description output (Table~\ref{tab:latency_results}).

\subsection{Performance of Hand Gesture and Gesture Recognition in Live Stream}\label{techEval_hand_gesture}
The goal of this evaluation was to assess the accuracy of our custom gesture recognition models in live video settings, beyond single-image performance. Unlike conventional model evaluations, we considered the combined performance of the recognition models and the temporal smoothing function (Section~\ref{keyframe_layer}). We reviewed gesture event logs and keyframes collected during the user study.

\subsubsection{Dataset and Analysis} 
During the user study, all keyframes and corresponding timestamps were automatically logged whenever a stable gesture state transition was detected. This enabled evaluation of both the gesture recognition models across sequences of frames and the effectiveness of the temporal smoothing algorithm. Each keyframe was labeled with a gesture state, including \emph{hold}, \emph{touch}, \emph{point}, and \emph{out of view}. In total, we collected 1,994 gesture instances, with 1,077 from the left hand and 917 from the right hand.

Each keyframe was manually annotated with a ground-truth gesture label by the research team. Gesture classes were assigned based on the visible hand pose, while the \emph{out of view} label was used when the wrist keypoint was not visible or when fingers were partially occluded by image boundaries, objects, or the other hand, conditions under which the Google MediaPipe hand landmark model~\cite{mediapipe} may fail.
We evaluated model performance by comparing predicted gestures to these ground-truth labels and computing standard metrics, including accuracy, precision, recall, F1-score, and confusion matrices for both hands.

\subsubsection{Results}
\begin{figure}[h]
\begin{center}
\vspace{-1pc}
\includegraphics[width=0.8\linewidth, alt={This is a confusion matrix heatmap for a 4-class classification model. Rows represent the actual (true) classes, and columns represent the predicted classes. Values are percentages. Higher numbers on the diagonal mean correct predictions. Matrix is presented below row by row (Actual vs. Predicted): Actual "Hold": Predicted as "Hold" = 75.5\%, "Touch" = 7.8\%, "Pointer" = 5.9\%, "Out of View" = 10.9\%. Actual "Touch": Predicted as "Hold" = 3.7\%, "Touch" = 88.0\%, "Pointer" = 3.7\%, "Out of View" = 4.6\%. Actual "Pointer": Predicted as "Hold" = 2.0\%, "Touch" = 6.9\%, "Pointer" = 55.9\%, "Out of View" = 35.3\%. Actual "Out of View": Predicted as "Hold" = 5.4\%, "Touch" = 5.7\%, "Pointer" = 5.0\%, "Out of View" = 83.9\%.}]{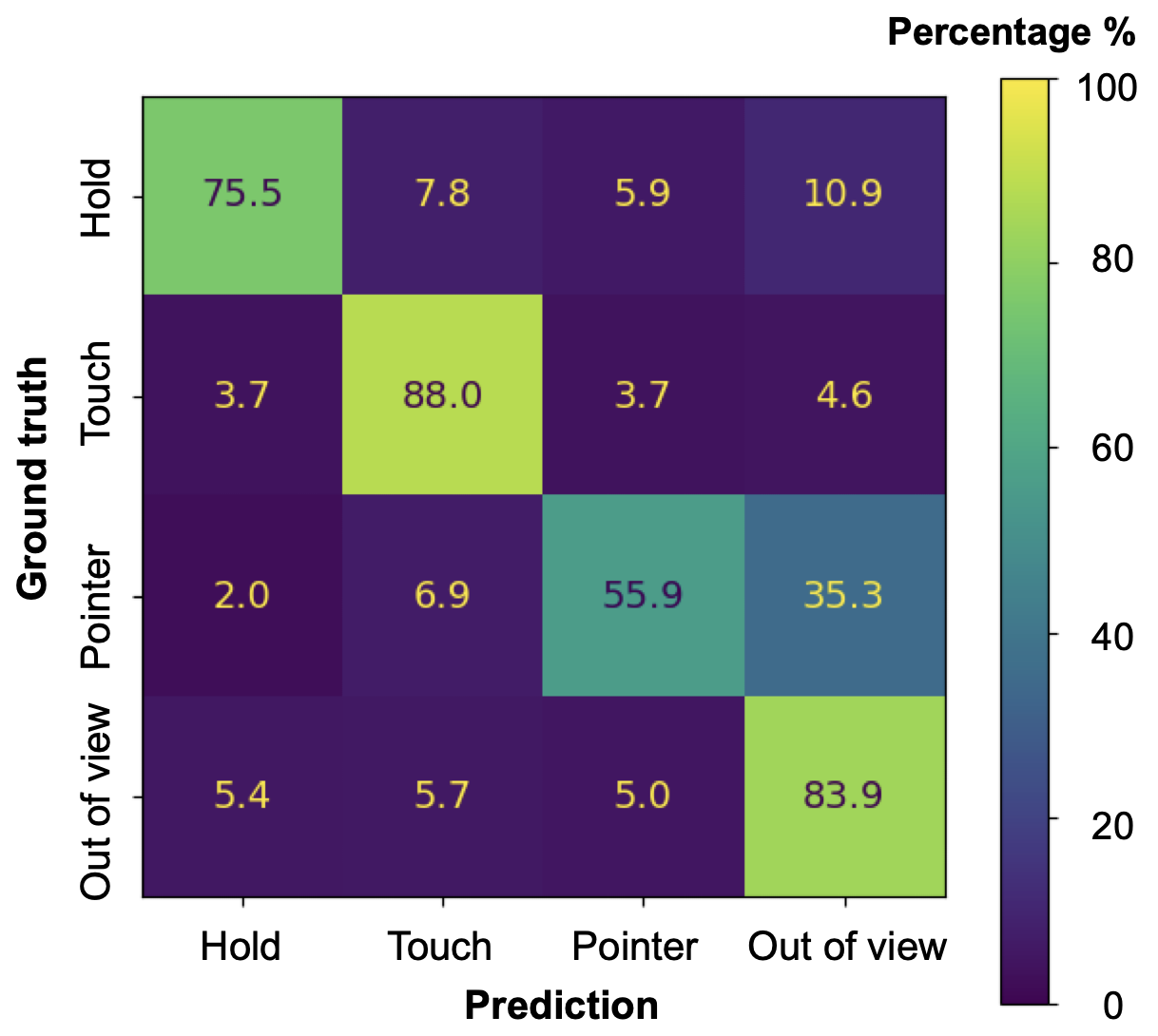}
\vspace{-1pc}
\caption{Confusion matrix for hand event detection of both hands.
}
\vspace{-0.5pc}
\label{fig:cm}
\Description{This is a confusion matrix heatmap for a 4-class classification model. Rows represent the actual (true) classes, and columns represent the predicted classes. Values are percentages. Higher numbers on the diagonal mean correct predictions. Matrix is presented below row by row (Actual vs. Predicted): Actual "Hold": Predicted as "Hold" = 75.5\%, "Touch" = 7.8\%, "Pointer" = 5.9\%, "Out of View" = 10.9\%. Actual "Touch": Predicted as "Hold" = 3.7\%, "Touch" = 88.0\%, "Pointer" = 3.7\%, "Out of View" = 4.6\%. Actual "Pointer": Predicted as "Hold" = 2.0\%, "Touch" = 6.9\%, "Pointer" = 55.9\%, "Out of View" = 35.3\%. Actual "Out of View": Predicted as "Hold" = 5.4\%, "Touch" = 5.7\%, "Pointer" = 5.0\%, "Out of View" = 83.9\%.}
\end{center}
\end{figure}

Among the 1,994 manually labeled instances, the model achieved an $F_1$-score of 0.77. 
Among the gesture classes (Figure~\ref{fig:cm}), the \emph{``hold''} gesture achieved the highest performance ($F_1 = 0.84$, precision $= 0.97$, recall $= 0.75$). 
The \emph{``touch''} gesture showed high recall ($0.87$) but lower precision ($0.60$), resulting in an $F_1$-score of $0.71$. 
Similarly, the \emph{``out of view''} class achieved an $F_1$-score of $0.74$ (precision $= 0.66$, recall $= 0.84$). 
The \emph{``point''} gesture, which had the fewest instances ($N = 102$), showed the lowest performance ($F_1 = 0.44$, precision $= 0.36$, recall $= 0.56$).

\begin{figure*}[h]
\begin{center}
\vspace{-0.4pc}
\includegraphics[width=\linewidth, alt={In image (a), a person holds two small bottles—one in each hand—while the green text label below reads “left: hold” and “right: hold.” In image (b), the person lightly touches two boxed items with both hands, and the green label reads “left: touch” and “right: touch.” In image (c), the person holds a small item in the left hand while pointing with the right, with green labels “left: hold” and “right: point.” In image (d), the left hand holds a spray bottle while the right hand touches the top, and the green labels read “left: hold” and “right: touch.” In image (e), left hand is holding a chocolate bar while right hand is away from the objects, and this time the labels are in red, reading “left: out” and “right: out.” In image (f), the left hand is holding a carton while the right hand touches it, with a red “left: out” label and a green “right: touch” label. In image (g), the left hand holds a small bottle and the right hand points at it, with green “left: hold” and red “right: touch” labels. In image (h), the left hand is not in view while the right hand moves along the edge, creating motion blur, with red “left: out” and red “right: touch” labels.}]{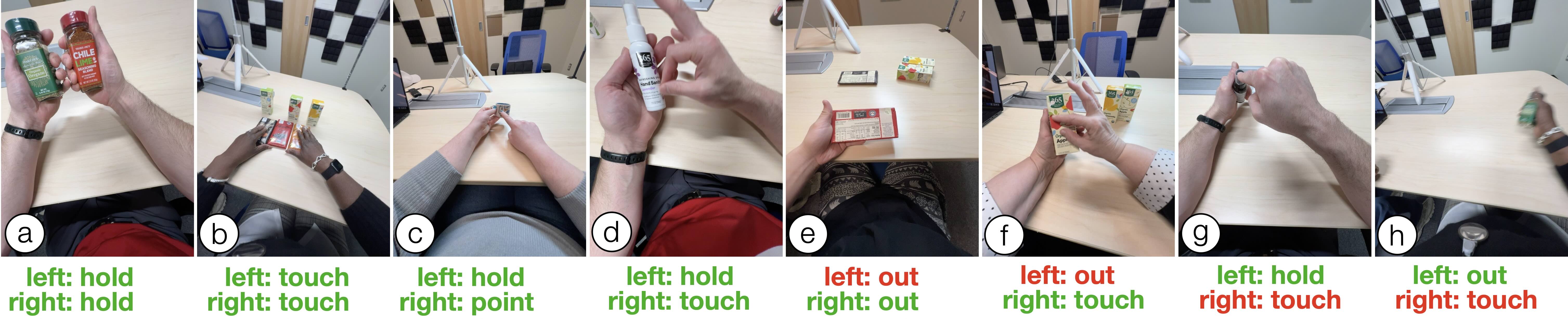}
\vspace{-1.7pc}
\caption{Example keyframes extracted by {\name} and the corresponding recognized gesture classes.
(a–d) {\name} successfully identified gestures from both hands across varying camera viewpoints.
(e) Recognition became challenging sometimes when hands were occluded, either by a larger object (e.g., a chocolate bar) or (f) by the user’s other hand during bimanual interactions.
(g) Gestures could also be misclassified under certain camera angles or hand postures, such as when the finger in a pointing gesture appears not extended from the camera’s viewpoint.
(h) Motion blur caused by camera or hand motion also influenced recognition reliability.
}
\vspace{-1pc}
\label{fig:camera_issues}
\Description{In image (a), a person holds two small bottles—one in each hand—while the green text label below reads “left: hold” and “right: hold.” In image (b), the person lightly touches two boxed items with both hands, and the green label reads “left: touch” and “right: touch.” In image (c), the person holds a small item in the left hand while pointing with the right, with green labels “left: hold” and “right: point.” In image (d), the left hand holds a spray bottle while the right hand touches the top, and the green labels read “left: hold” and “right: touch.” In image (e), left hand is holding a chocolate bar while right hand is away from the objects, and this time the labels are in red, reading “left: out” and “right: out.” In image (f), the left hand is holding a carton while the right hand touches it, with a red “left: out” label and a green “right: touch” label. In image (g), the left hand holds a small bottle and the right hand points at it, with green “left: hold” and red “right: touch” labels. In image (h), the left hand is not in view while the right hand moves along the edge, creating motion blur, with red “left: out” and red “right: touch” labels.}
\end{center}
\end{figure*}

False positives and negatives were observed under various conditions (Figure~\ref{fig:camera_issues}). 
For example, holding objects often resulted in partial or full hand occlusion, such as when grasping a juice carton or a box of chocolate bars (Figure~\ref{fig:camera_issues}e, f), which led to incorrect hand landmark detection and both types of errors. 
Similar occlusions occurred during discrete gestures like \emph{hold+point} and \emph{hold+swipe-up}. 
In addition, body movements and camera angles occasionally resulted in motion blurs and hands or fingers being partially cropped or outside the frame (Figure~\ref{fig:camera_issues}g, h). 
We discuss these camera-related issues and potential solutions in Section~\ref{discussion_camera}, as well as broader improvements to our vision-only approach in Section~\ref{discussion_sensor}.

\subsection{Latency of Delivered Descriptions}
Next, we measured the latency of descriptions to quantify how long users waited before feedback was read aloud. 
This included hand-state feedback, brief and detailed object descriptions, comparative descriptions, color labels, and object texts. We measured end-to-end latency as the time between detection of a new gesture and the onset of the corresponding spoken description.
This measurement encompassed the entire processing pipeline, including gesture recognition, retrieval of hand–object contact data and cropped images via Hands23~\cite{hand23}, prompt construction, response generation by VLMs (from Figure~\ref{fig:system_diagram}a to Figure~\ref{fig:system_diagram}c), and text-to-speech synthesis.

\subsubsection{Dataset and Analysis} In total, we analyzed all descriptions presented to participants during the study, comprising 1,143 instances of \emph{hand-state feedback}, 416 instances of \emph{brief object descriptions} generated by Moondream~\cite{moondream}, 208 instances of \emph{detailed object descriptions} generated by GPT-4o~\cite{gpt4o}, 35 instances of \emph{comparative descriptions} generated by GPT-4o, 529 instances of \emph{color labels}, and 143 instances of \emph{object texts}.
We measured the latency for each description type as the time between detecting a new gesture and sending the corresponding frame to VLMs, and the moment when the resulting description was delivered to the user. 
Below, we first report the processing time of individual components in the pipeline, followed by the end-to-end latency experienced by users.

\subsubsection{Results - latency of each model}
Because the latency of each component contributes to the overall end-to-end delay, we report individual model latencies to illustrate their respective performance (Table \ref{tab:latency_results}). Under the hardware configuration described in Section~\ref{implementation_details}, Hands23 exhibited an average latency of 0.87 seconds (SD=0.86), Moondream averaged 0.48 seconds (SD=0.62), and GPT-4o incurred the highest latency, with a mean of 3.07 seconds (SD=3.08).

\subsubsection{Results - end-to-end latency between gesture issued to descriptions delivered}
Under the hierarchical feedback design, \emph{brief object descriptions} typically followed \emph{hand-state feedback}, with detailed or comparative descriptions presented subsequently. In contrast, discrete gestures such as \emph{hold+point} for color identification and \emph{hold+swipe-up} for text retrieval allowed users to interrupt ongoing narration and quickly access targeted information (Section~\ref{responsiveness}).

In terms of end-to-end latency, \emph{hand-state feedback} exhibited a mean delay of 0.56 seconds (SD=0.91), providing near-immediate confirmation of system perception. Among all feedback types, color labels triggered by the \emph{hold+point} gesture had the lowest latency (M=0.09s, SD=0.17), while object text retrieval via \emph{hold+swipe-up} averaged 0.57 seconds (SD = 0.58).

In contrast, \emph{brief object descriptions} generated by Moondream averaged 5.36 seconds (SD=3.42), followed by \emph{detailed object descriptions} from GPT-4o with a mean latency of 10.3 seconds (SD=4.02). \emph{Comparative descriptions} from GPT-4o exhibited the highest latency, averaging 14.0 seconds (SD=3.06). Notably, these latency values account for the completion of prior descriptions, during which users were engaged with ongoing audio output rather than waiting idly.

\subsection{Accuracy of Object Descriptions from VLMs}\label{tech_eval_des_accuracy}
Lastly, we evaluated the accuracy of descriptions generated by VLMs. The goal was to determine whether the information presented to the user is accurate and relevant. 
For each referenced frame, we assessed whether {\name} correctly described the interacted object and whether any hallucinations appeared in the generated descriptions.

\subsubsection{Dataset and Analysis}
In total, we collected 802 descriptions from the study, including 416 \emph{brief object descriptions} from Moondream, 143 \emph{object texts}, 208 \emph{detailed object descriptions}, and 35 \emph{comparative descriptions} generated by GPT-4o. 
All instances were manually annotated for correctness. Descriptions were deemed incorrect if the system misidentified the interacted object or exhibited hallucinations.

\subsubsection{Results}
We evaluated the accuracy of 802 descriptions. 
\emph{Brief object descriptions} generated by Moondream achieved an accuracy of 91.59\% (381 out of 416). 
\emph{Detailed object descriptions} generated by GPT-4o reached 93.27\% accuracy (194 out of 208).
\emph{Comparative descriptions} generated by GPT-4o achieved 91.43\% accuracy (32 out of 35). 
Overall, the descriptions demonstrated strong accuracy, with common errors including misidentifying a chocolate bar as a book, hallucinating a mouse when users rested their hands on the table, referencing the table instead of the held object, or failing to describe objects when they were occluded by hands.
In contrast, \emph{object texts} achieved lower accuracy at 67.83\% (97 out of 143). This was primarily due to challenges in recognizing text on the cylindrical bottles used in the study, where curvature often distorted or partially occluded the text. Participants frequently relied on trial-and-error repositioning to present readable text to the camera. We discuss potential mitigation strategies in Section~\ref{discussion_camera}.
\section{Discussion and Future Work}\label{discussion}
We discuss our lessons learned and design implications for generating live object descriptions with hands as natural information cursors.

\subsection{Supporting Customization and Adaptation of Broader Gesture Set}\label{discussion_customization}
To our knowledge, {\name} is the first system to deliver live, rich descriptions driven by diverse hand–object interactions. {\name} requires users to learn a predefined gesture set. This set, though informed by prior research on \emph{gesture nature} and \emph{BLV familiarity} (Section~\ref{gesture_set}), resulted in perceived cognitive load by our participants (Section~\ref{results_cognitive_load}). Nevertheless, qualitative feedback highlighted the utility of gestures and the enhanced sense of agency they afforded in accessing object information (Section~\ref{results_intuitiveness}).

The current gesture set serves as a foundation that can be extended through user customization~\cite{objecttrain, Oh2013, GestureCustomizer, TouchCam, EarTouch}. Such flexibility is essential given the diversity of gesture preferences and contextual needs among BLV users. Social acceptability, in particular, could play a key role in shaping gesture choice. For example, mid-air gestures may be more suitable in private settings, where BLV users have been observed performing metaphoric gestures for tasks such as TV control, while sighted users tend to employ symbolic gestures~\cite{TVgestures}. In contrast, subtle micro-gestures or touchscreen interactions are often preferred in public environments due to their discreet nature~\cite{Oh2013, EarTouch}.

Accordingly, while {\name} currently incorporates two micro-gesture interactions (e.g., pointing to a held object for colors and swiping up for available texts) to demonstrate feasibility, the gesture vocabulary could be expanded by drawing on interaction techniques from existing assistive technologies, such as touchscreen screen readers.
Examples include swiping left or right to navigate at the word level, using multi-finger swipes to access higher-level semantic information, and familiar interactions such as pinch-to-zoom for localized text exploration.

Future work could involve systematic elicitation studies with BLV users to capture gesture preferences across public and private contexts, as well as the development of adaptive AI companions capable of learning and personalizing gesture mappings over time.

\subsection{Design Implications for Low-Latency, Context-Aware Gesture Recognition}\label{discussion_sensor}
Hand movements are inherently complex and dynamic, making them difficult to capture reliably using a camera stream alone. 
Participants observed occasional interruptions in descriptions while using {\name} (Section~\ref{results_intuitiveness}), which were attributable to limitations in our custom gesture recognition models (Section~\ref{techEval_hand_gesture}). Even when at rest, hands may unintentionally resemble supported gestures. These erroneously detected gestures prompted {\name} to start generating new descriptions.

Incorporating additional contextual cues could be essential to mitigate unintended gesture recognition by better distinguishing intentional from unintentional hand activities. 
For example, high-level user activities can be inferred from full-body posture~\cite{MeCap} and enriched through on-body sensors or wearable devices~\cite{WheelPoser, IMUPoser, PoseOnTheGo, Khanna2024, GestureCustomizer}, enabling the system to disregard situations in which hands are merely resting on objects or laps, or casually moving during locomotion.
Furthermore, object contact and categories may be inferred from complementary sensing modalities, such as acoustic signals~\cite{SAWSense} and electromyography~\cite{Fan2018}.
Incorporating diverse sensing techniques could help reduce reliance on a vision-only pipeline, particularly susceptible to occlusion, and support cross-validation of gesture and object recognition across modalities (Section~\ref{techEval_hand_gesture}).

Beyond sensor fusion, embedding common knowledge about everyday object use, typical hand postures, and users’ habitual interaction patterns could further filter out irrelevant contexts, such as hands resting on tables or interacting with familiar items like keyboards, laptops, or mice.
Additionally, inaccuracies in text recognition arising from curved surfaces (Section~\ref{tech_eval_des_accuracy}) could be alleviated by recognizing and combining texts from multiple previously captured views of an object’s surface.

Recent advances in VLMs, including improvements in both accuracy and latency and the emergence of lightweight models such as Gemini-Flash~\cite{geminiflash}, suggest that system responsiveness will continue to improve. Reduced latency also allows greater temporal budget for incorporating these complementary sensing components into the description pipeline, which could further enhance overall system reliability and accuracy.

\subsection{Trade Offs between Camera Devices, Configurations and Practicality}\label{discussion_camera}
Camera-based ATs face several long-standing challenges~\cite{Aaron2012, LastMeter, Lee2019ASSETS, HandsHoldingClues, Guo2018Cursor, Manaswi2019, VisPhoto, zquez2012}, including maintaining target objects within the camera frame~\cite{Guo2018Cursor, zquez2012}, ensuring adequate coverage of essential visual content~\cite{LastMeter, VisPhoto}, and addressing the social acceptability of camera setups~\cite{Koelle2019, Profita2016, Akter2022}.
While these considerations informed the design of {\name} (Sections~\ref{gesture_set} and~\ref{implementation_details}), further work is needed to support practical deployment in real-world contexts.

In {\name}, we employed a neck-mounted smartphone to free users’ hands and approximate an egocentric perspective. This design was inspired by the potential of emerging smart glasses, which at the time of development involved several trade-offs. Smartphones, by contrast, offered more accessible APIs than commercial smart glasses, greater flexibility in adjusting camera resolution and FoV (e.g., standard versus wide-angle), and sufficient battery life to support extended study sessions. Although this setup met our research needs, neck-mounted cameras differ from head-mounted configurations, requiring additional synchronization between head orientation and hand movements~\cite{han2024wearables, hersh2022wearable}. Moreover, such setups may be uncomfortable for prolonged use or raise concerns regarding social acceptability in everyday contexts~\cite{Profita2016, aisee, Akter2022}, especially given varying privacy sensitivities among BLV users and bystanders.

We selected a wide FoV (FoV; 120°) rather than a standard FoV (77°) to balance coverage and distortion. While standard FoV lenses introduce minimal distortion and support more reliable hand detection, their limited coverage makes it difficult to capture both hands and relevant objects simultaneously. Consequently, we opted for a wide-angle lens to increase coverage despite its greater distortion, which negatively affected hand detection performance (Section~\ref{techEval_hand_gesture}).
Although {\name} provided feedback on perceived hand states, participants were often unaware of the camera’s intrinsic limitations, as reflected in comments such as: \emph{``I’m wondering, does closer to the camera matter?} (P2) and \emph{``I’m blind, so I don’t think about how the camera looks and stuff. So this is all good learning.''} (P5). 

Building on this feedback, future research could explore a broader range of camera-mounting configurations (e.g., body-mounted or head-mounted) to better accommodate individual preferences, comfort, and social contexts. This may involve integrating additional sensors, such as IMUs in wrist- or head-mounted devices, and providing feedback to address head–hand misalignment, such as haptic–audio guidance techniques for camera aiming~\cite{Jayant2011, Adam2013}. Adaptive lens-selection strategies based on hand–camera distance and object distribution could further improve coverage and accuracy; for example, switching to a wide FoV to cover multiple objects and to a standard FoV when focusing on a single object.

\subsection{Gesture-driven Descriptions Beyond Physical Reach}
{\name} delivers live visual descriptions driven by hand–object interactions within reach. Participants found this approach intuitive (Section~\ref{results_intuitiveness}) and more efficient than photo-capturing and chat-based interactions in current assistive apps.
While {\name} centers on holding and touching objects, this may not always be feasible due to social stigma, safety concerns, or personal comfort. We discuss circumstances that limit tactile engagement, and outline potential ways to support gesture-based interaction even when physical reach is constrained.

Cultural taboos surrounding public touch, reinforced by norms such as the ubiquitous museum rule of ``don’t touch'', can lead BLV individuals to internalize tactile exploration as socially inappropriate~\cite{donttouch}. 
Additionally, some BLV individuals may avoid touch due to negative prior experiences, such as being compelled to explore unfamiliar objects without preparation, consent, or agency~\cite{stigma, mclinden2016learning}. 
Beyond social stigma, tactile exploration can also present safety concerns, especially during public health crises such as the COVID-19 pandemic~\cite{covid19, 2covid19, alves2023living}. 
These challenges are further compounded by physical constraints, as some objects of interest, such as items placed on high shelves in grocery stores, may be inaccessible. 

To extend gesture-driven descriptions beyond direct physical reach, future systems could build upon prior work on interaction proxies~\cite{InteractionProxies, HandProxy, Tao2024} and camera motion–enabled live description tools~\cite{worldscribe, seeingai} (Table~\ref{tab:cursors}). 
For example, after receiving an initial overview of a visual scene and confirming interest, users could employ subtle mid-air gestures~\cite{visionprogesture, Khanna2024} (e.g., pinch) or touch gestures on an interaction proxy (e.g., touchscreen) to navigate details with audio feedback.
Such integrations would broaden access to visual environments both within and beyond physical reach.

\section{Conclusion}
In this work, we introduced {\name}, a system that augments hand-object interactions with automated, live visual descriptions. 
By leveraging egocentric hand gestures as information cursors, {\name} enables users to enrich their understanding of objects through diverse interaction patterns, such as holding or touching an object to receive hierarchical descriptions, comparing objects by holding them side by side, and swiping upward to read available text.
Through a controlled user study and technical evaluation, we demonstrated that {\name} delivers reasonably accurate, timely, and informative feedback to support BLV users across a range of object exploration tasks.
Participants perceived {\name} to be easy to learn and intuitive, and felt in control when accessing object information. 
Finally, we discussed implications for real-world deployment, including supporting gesture and information customization, improving gesture recognition and description accuracy through broader contextual awareness, considering diverse camera configurations and social acceptability, and extending hand-driven interaction beyond physical reach.

\begin{acks}
We thank our anonymous reviewers and all the participants in our study for their feedback and suggestions. Ruei-Che Chang was supported by the Apple Scholars in AI/ML PhD fellowship. Rosiana Natalie was partially supported by the Michigan Data Science Fellowship at the University of Michigan. This research was also supported in part by a Google Academic Research Award and a Google Cloud Platform Credit Award. 
\end{acks}

\bibliographystyle{ACM-Reference-Format}
\bibliography{touchscribe}

\onecolumn
\appendix

\section{Tables}
\begin{table*}[h]
  \caption{Participant demographic information, referred to as P1 to P8.}
  \label{tab:demographic}
  \vspace{-1pc}
  \begin{center}
  \begin{tabular}{|l|l|l|p{7cm}|p{4.5cm}|}
    \hline
    \textbf{ID} & \textbf{Age} & \textbf{Gender} & \textbf{Self-Reported Visual Ability} & \textbf{Assistive App Use} \\
    \hline
    P1 & 41 & Male & Blind due to Retinitis Pigmentosa, left < 0.5 degree, depends on lighting to identify the color of the object. & SeeingAI, BeMyAI, BeMyEyes, Aira, Orcam, SoundScape, and VoiceVista \\
    \hline
    P2 & 58 & Female & Right: blind. Left: Usable vision using a physical magnifier. & SeeingAI, BeMyAI, BeMyEyes, Aira, and Orcam, \\
    \hline
    P3 & 50 & Female & Blind, since birth. Light perception. & SeeingAI, BeMyAI, BeMyEyes, Aira, Orcam and BlindSquare \\
    \hline
    P4 & 73 & Female & Blind, since birth. Light perception. & SeeingAI, BeMyAI, BeMyEyes, and Aira \\
    \hline
    P5 & 41 & Male & Blind, since birth. Light perception. & SeeingAI, BeMyAI, BeMyEyes, and SoundScape\\
    \hline
    P6 & 60 & Female & Blind, since birth. & BeMyAI and BeMyEyes \\
    \hline
    P7 & 24 & Female & Blind, acquired since 13. & None \\
    \hline
    P8 & 18 & Male & Low vision due to Stargardt. Right: 20/1000, Left: 20/600, Light to Moderate color blindness. & SeeingAI \\
    \hline
  \end{tabular}
  \end{center}
  \Description{}
\end{table*}

\begin{table*}[h]
\centering
\caption{Accuracy of object descriptions generated by VLMs.}
\vspace{-1pc}
\label{tab:accuracy_results}
\begin{tabular}{lccc}
\toprule
\textbf{Description Type} & \textbf{Instances} & \textbf{Correct} & \textbf{Accuracy (\%)} \\
\midrule
Object labels (Moondream) & 416 & 381 & 91.59 \\
Detailed descriptions (GPT-4o) & 208 & 194 & 93.27 \\
Comparative descriptions (GPT-4o) & 35 & 32 & 91.43 \\
Object texts (GPT-4o) & 143 & 97 & 67.83 \\
\bottomrule
\end{tabular}
\end{table*}

\begin{table*}[h]
\centering
\caption{Latency results for description generation, reported as mean and standard deviation (SD) in seconds.}
\vspace{-1pc}
\label{tab:latency_results}
\begin{tabular}{lccc}
\toprule
\textbf{Description Type / Model} & \textbf{Instances} & \textbf{Avg. Latency (s)} & \textbf{SD (s)} \\
\midrule
\multicolumn{4}{c}{\emph{Model Processing Latency}} \\
\midrule
Hands23~\cite{hand23} & -- & 0.87 & 0.86 \\
Moondream~\cite{moondream} & -- & 0.48 & 0.62 \\
GPT-4o~\cite{gpt4o} & -- & 3.07 & 3.08 \\
\midrule
\multicolumn{4}{c}{\emph{End-to-End Latency (Gesture Identified $\rightarrow$ Description Spoken)}} \\
\midrule
Hand state descriptions & 1143 & 0.561 & 0.91 \\
Object labels (Moondream) & 416 & 5.36 & 3.42 \\
Detailed descriptions (GPT-4o) & 208 & 10.3 & 4.02 \\
Comparative descriptions (GPT-4o) & 35 & 14.0 & 3.06 \\
Object texts (GPT-4o) & 143 & 0.57 & 0.58 \\
Color labels & 529 & 0.087 & 0.169 \\
\bottomrule
\end{tabular}
\end{table*}

\end{document}